# Development of Interatomic Potentials to Model the Interfacial Heat Transport of Ge/GaAs

Spencer Wyant[1], Andrew Rohskopf[2], Asegun Henry[2]

[1]Department of Materials Science and Engineering, Massachusetts Institute of Technology, 77 Massachusetts Avenue, Cambridge, Massachusetts 02139, USA

[2] Department of Mechanical Engineering, Massachusetts Institute of Technology, 77 Massachusetts Avenue Cambridge, Massachusetts 02139, USA

## Abstract

Molecular dynamics simulations provide a versatile framework to study interfacial heat transport, but their accuracy remains limited by the accuracy of available interatomic potentials. In the past, researchers have adopted the use of analytic potentials and simple mixing rules to model interfacial systems, with minimal justification for their use. On the other hand, contemporary machine learned interatomic potentials have greater complexity, but have not seen rigorous validation of interfacial heat transport properties. Moreover, when fitting to *ab initio* data, it is not known whether interface systems small enough to be tractable for density functional theory calculations can produce reasonable interatomic force constants. These and related questions are studied herein using a model Ge/GaAs system, with a particular focus on the harmonic force constants (IFC2s) of the interface. The *ab initio* IFC2s are shown to recover near bulk-like values ~ 1-2 nm away from the interface, while also exhibiting a complex relationship across the interface that likely precludes any successful application of mixing rules. Two different spectral neighborhood analysis potentials (SNAP) are developed to model the interface. One is fit to the total forces, while the other is only used to describe the anharmonicity, with a Taylor expansion used to describe the harmonic portion of the potential. Each potential, along with their merits and issues are compared and discussed, which provides important insights for future work.

## I. Introduction

The physics of thermal boundary conductance (TBC) is not well understood, yet it plays an increasingly important role in the thermal management of nanoscale devices [1–4] and can be leveraged to enhance performance in applications like thermoelectrics [5,6] and thermal barrier coatings [7,8]. Experimental efforts to accurately measure the TBC of interfaces have expanded tremendously over the past couple decades [9–11]. However, challenges remain in developing sufficiently sensitive metrology for some material systems and in de-convoluting the effects of structural and compositional disorder, strain, and defects near interfaces.

Theoretical efforts are an essential complement to experiments, with the ability to clarify the underlying physics and to enable rational design by systematically exploring the effects of different interfaces and interfacial heterogeneities. One class of theoretical approaches that has been extensively used to try to understand TBC is based on the phonon gas model (PGM) and the Landauer formula, which expresses conductance in terms of phonon particles carrying energy through an interface with some probability, termed the transmission probability [12]. A couple of simple approximations have historically been used to model this transmission probability, namely the acoustic mismatch model [13,14] and diffuse mismatch model [15,16], and recently more sophisticated atomistic-level approaches like the atomistic green's function (AGF) [17–19] and wave packet method [20,21] have been employed.

Aside from AGF, it is also possible to directly simulate the atomistic dynamics of a system using molecular dynamics. Such an approach is valuable in its generality, since it is capable of studying arbitrarily complicated systems, with full inclusion of temperature dependent anharmonicity. Additionally, MD enables the study of both statistically averaged properties and detailed temporal dynamics, albeit in a classical limit. With respect to predicting TBC values, interfaces can be simulated under an equilibrium or nonequilibrium framework [22,23], and methods like interface conductance modal analysis (ICMA) [24–26] can be used to determine TBC and provide a detailed analysis of the modal contributions.

At the heart of an MD simulation is the choice of the interatomic potential, which models the potential energy surface (PES) of the system and the resulting forces experienced by atoms as a function of their respective positions. Aside from the limitations associated with simulating classical dynamics (i.e., not accounting for quantum dynamics or statistics), the accuracy of an interatomic potential is the primary determinant of the accuracy of an MD simulation and any properties extracted from it. Specifically, in regards to thermal properties, the key values of interest are the interatomic force constants (IFCs), which can be obtained by taking a Taylor expansion (TE) of the PES at a given configurational minima. The lattice dynamics formalism, used to define and compute many thermal properties, generally takes as a primary input the second order (harmonic) and higher order (anharmonic) IFCs of a system [27]. Harmonic IFCs (IFC2s) determine the system's eigenmodes while the anharmonic IFCs govern the interactions between the modes. Developing interatomic potentials that accurately replicate the IFCs relative to some trusted reference is therefore essential for the reliable prediction of thermal properties using MD simulations.

In the past decade, a number of researchers have developed potentials focused on accurately modeling the thermal properties of bulk systems. Generally, *ab initio* data has been used as a reference, since methods like density functional theory (DFT) have been highly successful in predicting the phonon dispersions and thermal conductivity of bulk systems, the latter via a Boltzmann Transport Equation (BTE) framework [28]. One approach has been to take common analytic potentials and fit them to *ab initio* forces and IFCs [29–33], although high accuracy with respect to thermal properties can be difficult to achieve due to the simple functional form of these potentials [34]. In contrast, machine learned interatomic potentials (MLIPs) utilize highly flexible functional forms that can reproduce an *ab initio* PES with high fidelity, though typically requiring an appreciable amount of training data. The bulk thermal properties of a wide variety of materials systems have been successfully modelled with MLIPs based on linear models [35–37], Gaussian processes [38–40], and neural networks [41–46].

Another approach to modeling bulk systems with high thermal accuracy is to leverage the definition of IFCs and use a Taylor expansion directly as the interatomic potential. Such Taylor expansion potentials (TEPs) by definition have exact force constants if parameterized by the *ab* *initio* values and as a result can achieve extremely high force accuracy when incorporating 3$^{rd}$, 4$^{th}$ and higher order terms [47–50]. Nevertheless, their widespread adoption among the thermal transport community has been hindered by frequent instability problems [34,50]. To overcome this, Rohskopf et al. recently developed a method combining a harmonic translationally-invariant TEP (TITEP) with an additional analytic potential or MLIP [51], obtaining high force accuracy, exact harmonic IFCs, and good 3$^{rd}$ order IFC accuracy.

While the development of "vibrationally accurate" interatomic potentials to model bulk systems has seen considerable progress, the situation with interfaces remains relatively unexplored, with unique challenges compared to modeling bulk systems. Such an interatomic potential needs to accurately model three distinct regions -- the two bulk regions away from the interface and the interfacial region itself. The interface is especially challenging, as it exhibits a unique bonding environment where atoms from the respective bulk constituents interact with each other in a chemically distinct way. Here, a fundamental choice needs to be made as to how to handle these bonds. Should the interaction be modeled in a way that explicitly incorporates interface data during fitting, or can a simpler approach be taken, whereby interactions are derived purely from constituent bulk interactions?

Historically, researchers studying the thermal properties of interfaces using MD methods have adopted this latter approach, often out of a lack of better/easier options. The simplest of such approaches is to use a single potential to model both bulk regions and only vary the mass between the two sides [52–54]. A somewhat more versatile approach that has seen widespread adoption is the use of mixing rules to model interface bonds, in which the parameters of two bulk potentials of the same functional form are mixed according to specific rules. Despite being widely used in the past for many interface systems [55–61], it is unclear how well potentials with mixed parameters can represent the bonds across an interface, and specifically, their associated force constants.

If a mixing rule strategy cannot capture interface bonds well, it is necessary to develop a potential fit explicitly to interface data (in addition to bulk data). Here one of the most viable strategies is to use a MLIP, leveraging the flexibility of their functional forms to model the different bonding environments of bulk and interface regions with a single potential. Another interesting route would be the development of a TE-style potential, promising near exact IFCs if a method is developed to incorporate bulk and interface IFC sets into a single potential. While a number of prior studies have used MLIPs to model interfaces [41,62–66], only one was explicitly focused on interfacial thermal transport . Likewise, to the best of our knowledge, only a single paper has used a TEP to model an interface system, and it did not incorporate explicit *ab initio* interface IFCs [67].

However, a couple key questions need to be addressed before we can confidently use MLIPs and TE-style potentials fit to *ab initio* data to study thermal transport at interfaces. For one, it is not clear how large the DFT supercell needs to be in order to model an interface, so that its underlying force constants transition back towards their bulk values. Secondly, it is rarely, if ever, assessed as to how well a given MLIP can approximate the force constants at an interface region, and even then, the IFCs of bulk regions are usually only evaluated indirectly via dispersion and thermal conductivity calculations. Given their importance to thermal transport, the ability of MLIPs and related potentials to reproduce interface IFCs needs to be explicitly evaluated.

In this study, we answer the above questions through an investigation of a Ge/GaAs interface as a model system. Ge/GaAs is specifically of interest because it is nearly lattice matched, rendering it straightforward to study with MD (i.e., minimal defects/dislocations). In addition, the TBC of epitaxial Ge/GaAs samples can be measured using a novel transducerless thermoreflectance techniques [68].

To start, we obtain *ab initio* harmonic IFCs for the Ge/GaAs interface system and examine the convergence of the IFC2s with distance from the interface. These interface force constants are then used to assess whether mixing rules can be an effective model for the interfacial bonds. We then turn our focus to methods that explicitly model interface interactions and evaluate their performance with respect to bulk and interface thermal properties. We study a simple MLIP, a spectral neighborhood analysis potential (SNAP) [69], and a hybrid potential inspired by the work of Rohskopf et al. [51], which combines bulk and interface TITEPs with a single anharmonic SNAP potential, directly incorporating IFC2s into a potential that models an interface system.

## II. Convergence of Interface IFC2s Towards Their Bulk Values

When studying the properties of a single interface, we are typically interested in simulating the semi-infinite case in which an interface exists in isolation, formed by two infinite bulk materials coming into contact. Experimentally, this approximately corresponds to an atomically planar interface with many nanometers, if not micrometers, of bulk material separating it from the nearest boundary. When using classical molecular dynamics, relatively large systems with thousands of atoms can be employed, due to the linear scaling with the number of atoms, which is typically sufficient to realize convergence to the semi-infinite case.

However, when developing an interatomic potential fit to *ab initio* data, the size of the corresponding *ab initio* interface model has to be considered. Due to the roughly cubic scaling of most DFT methods, only small interface models (< 500 atoms) can be treated. There are two primary consequences of this. First the smaller system size constrains the extent to which the interface can relax, resulting in a different interface region than would be observed in the true semi-infinite case. Second, the force an atom experiences can be affected by atoms both within the cell and by periodic images of those atoms, when using periodic boundary conditions.

In turn, this means that the structural model and resulting forces the interatomic potential is fit to, differ from the ideal semi-infinite case, effectively sampling a different region of the potential energy surface. In regards to thermal properties, the underlying interatomic force constants will differ, partly as a function of the different structural model, and partly due to the incomplete evolution of IFCs towards their bulk values. When the resulting potential is used in a larger system, the potential has to effectively infer this full evolution and will necessarily need to "stitch" the underlying force constants of the bulk and the small interface model. For a TEP, this stitching is explicit, as the force constants and equilibrium structure are hard-coded into the potential; for a MLIP, this stitching is implicit, as the potential will relax the large interface structure according to its learned forces from bulk structures and the smaller DFT interface.

Focusing on the structural convergence in the out-of-plane direction, in the ideal case the DFT interface model needs to be sufficiently long to allow for nearly complete relaxation of the interface and the recovery of near bulk-like IFC2s away from the interface. As a consequence, minimal extrapolation would be needed for the fitted potential to simulate a large interface model. Additionally, the convergence of longer-range force constants can provide a convenient proxy for the overall structural convergence of the interface system. However, it is unclear if tractable DFT-sized interface models can in fact satisfy these conditions.

To study this question, we use DFT to relax and obtain force constants for bulk Ge, GaAs and a Ge-GaAs interface model. Plane-wave DFT calculations are performed using the Vienna Ab Initio Simulation Program (VASP) [70–72] under the local density approximation (LDA). Full details of the DFT calculations are provided in the Supplementary Information (SI).

The Ge/GaAs interface is modelled in a superlattice geometry rather than a slab geometry, as the vacuum region associated with slab calculations considerably increases their computational expense, limiting the generation of *ab initio* fitting data for potential fitting discussed in later sections. The interface is composed of the equivalent of a 2x2x7 supercell of the conventional unit cell of bulk Ge/GaAs. The GaAs side of the interface is strained in-plane to match the Ge lattice parameters, since corresponding experimental samples were fabricated with a thin GaAs layer that conforms to the underlying Ge single crystal substrate (presented in a future work). Correspondingly, the bulk GaAs is also strained in-plane to Ge lattice parameters (henceforth referred to as s-GaAs) and allowed to relax, resulting in a small tetragonal compression to compensate for the expansion of the in-plane lattice parameter.

The interface itself is a sharp planar As-terminated interface with no atomic intermixing. While a real interface should involve some level of atomistic intermixing to compensate for the Coulombic imbalance [73], here we have explicitly chosen an unmixed interface, to simplify our assessment of IFC2s in this section and the next. Future work, however, should consider intermixing to achieve neutrality. To ensure that both interfaces are identical and increase the symmetry of the system, a nonstoichiometric model is needed so as to create an As-terminated interface at both interfaces.

The interface superlattice is relaxed using a procedure similar to that used recently by Lindsay and Polanco [74]. First, the out-of-plane lattice z parameter and the z component of the atomic coordinates are simultaneously minimized. Then with the lattice vectors fixed, all atoms are allowed to relax with full atomic degrees of freedom. The resulting structure, seen in Figure 1a, exhibited minimal distortion. The plane separation distance along the out-of-plane z-direction is within 0.32% and 1.4% of their respective bulk Ge/s-GaAs values one plane from the interface, and eventually settle to 0.23%/0.17% of the respective bulk values at the midplanes (see Fig. S2 of the SI). There was no displacement in the x-y direction, even with symmetry preservation turned off during relaxation.

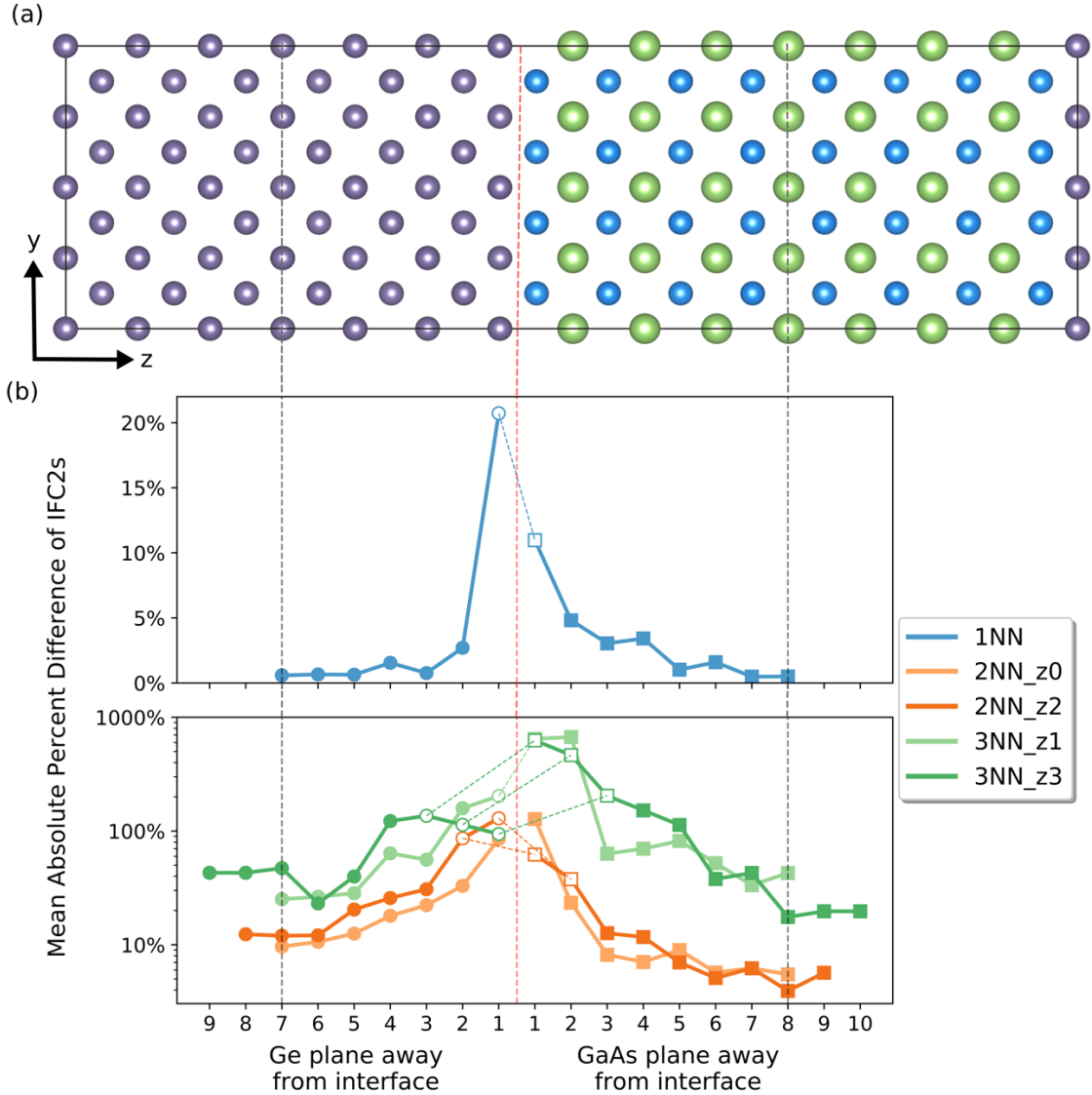

*Figure 1: (a) Relaxed Ge/GaAs interface system used in DFT calculations. Red dashed line indicates the interfacial plane, while grey dashed lines indicate midplanes between the two interfaces of the system. (b) Convergence of the IFC2s with the length of the system. Mean absolute percent differences of each type of IFC2 tensor relative to its respective bulk values are plotted for each plane moving away from the interface. On the Ge side, IFC2s are indexed by the left atom (lower z) of the atomic pairs associated with the IFC2s. On the GaAs side, IFC2s are indexed by the right atom (higher z). IFC2s corresponding to atom pairs that cross the interfacial plane, represented by open symbols, appear twice on this plot, once normalized to bulk Ge values and once normalized to bulk s-GaAs. As a visual aid, a thin dashed line connects these equivalent IFC2 sets. Note that the lower plot is plotted on a log scale, while the upper plot is not.*

To obtain harmonic force constants, we used the small finite displacement approach, as implemented in the freely available code for calculating phonon properties – Alamode [75]. To aid in the tractability of the problem, we limited the cutoff of these force constants to 5.715 Å, corresponding to 3NN interactions. Overall, the high symmetry of the system is reflected in the relatively low number of unique force constants for the interface system, with 507 total 1NN, 2NN and 3NN IFCs. This is in stark contrast to, for example, an idealized Al/Al2O3 interface which can have >5,000 IFCs with a similar sized cutoff [76]. Moreover, the force constants at both the middle and periodic boundary interface are identical, and as such we only need to consider IFC2s from the interface to each midplane.

The convergence of IFC2s with respect to the out-of-plane length of the system is shown in Figure 1b. IFC2s are distinguished by the character of the atomic pairs they are associated with, namely their nearest neighbor bin and the number of atomic planes spanned in the out-of-plane z direction. For example, IFCs labeled by "3NN_z3" correspond to third nearest neighbor atom pairs spanning 3 planes in the z direction, while "2NN_z0" correspond to second nearest neighbor atom pairs located within the same z plane.

Broadly speaking, these plots demonstrates that there is appreciable IFC2 convergence towards their respective bulk values on both the Ge and GaAs sides of the system within 1-2 nm. 1NN IFC2s, the largest harmonic force constants by an order of magnitude, are already 2.7% and 4.8% of their respective bulk Ge and bulk s-GaAs values only two planes away from the interface (i.e., the first 1NN IFC2s associated with atom pairs not crossing the interfacial plane). These 1NN IFC2 values eventually settle to 0.59% and 0.49% of the respective bulk Ge and bulk s-GaAs values at the midplane. The 2NN IFC2s, while not as strongly converged, still settle between 5-13% of their respective bulk values at the midplanes. Unsurprisingly the 3NN IFC2s, which are roughly 75x smaller than 1NN IFC2s and can span up to 3 planes in the z direction, are the least converged. But even these IFC2s settle to within 25-44% of their respective bulk values.

The observed convergence provides the first explicit confirmation that tractable DFT interfaces can exhibit a nearly complete transition back to bulk behavior with respect to the 1NN IFCs. The convergence of 2NN and 3NN IFCs also appear reasonable, though they are an order of magnitude less converged than the 1NN IFCs, despite the minimal distortions induced by relaxation. This demonstrates the way in which IFC2 convergence can be considered a more stringent test for overall PES convergence, as longer-range IFC2s by their very nature will reflect longer-range structural features and remnants of electronic structure features influenced by the presence of the interface and the other material. By assessing how well the IFC2s converge back towards their bulk values, we are by proxy assessing how close our DFT model can directly approximate the "true" PES of the semi-infinite case. Clearly, even for a high symmetry system with minimal distortion, this is still a challenging goal, but it is tenable with today's computing power.

What would the implications be for DFT interface systems whose IFC2s exhibit poorer convergence than demonstrated here? For a TE-style potential, when IFC2 values are well converged, the explicit stitching between bulk and interface regions needed to produce a large hessian for MD simulations is a simple task, since it would be justified to use the bulk IFC2 values in those regions. However, if interface IFC2s were less converged in these regions, the use of bulk IFC2 values would introduce large discontinuities in the force constant profile, perhaps resulting in artifacts in the system's modes and instabilities. Meanwhile, for an MLIP fit to a poorly converged interface system, the potential may have to operate in a somewhat extrapolative regime in order to relax a large interface model. While it may still be able to obtain a seemingly sensible relaxation profile, structurally and in terms of IFC2s, there would be limited ability to validate such a result without a more converged *ab initio* reference.

## III. Assessing the Applicability of Mixing Rules

The use of mixing rules to combine the parameters of interatomic potentials was initially deployed to model fluid mixtures of noble gases and organic molecules [77,78]. Beginning with simple Lorentz-Bertholet rules (arithmetic and geometric means) applied to the Lennard-Jones potential, a significant amount of scholarship has been spent iteratively developing better mixing rules to model these mixtures [79–85], often with a level of physical justification, and evaluating their performance in comparison to experimental results [78,86–90]. Over time, mixing rules were applied to model non-Van der Waals solid alloys or interfaces using more complex interatomic potentials like Stillinger-Weber [91] or Tersoff [92]. However, only the simplest Lorentz-Bertholet rules have been used (though an additional parameter is available for Tersoff to adjust interactions), with minimal physical justification and only limited validation focused primarily on bulk energetic properties [93–97].

Ultimately the use of these simple mixing rules was borne out of necessity, as researchers interested in studying alloys and interfaces were constrained computationally, unable to perform the large and/or numerous *ab initio* calculations that would be needed to validate and guide the development of better potentials. Moreover, modern MLIP methods had yet to be developed even if such *ab initio* data were available, and so it would have been challenging to translate this accurate data into an interatomic potential. Since these large *ab initio* calculations are much more tractable now, we can use well converged IFC2 sets like those discussed in the prior section to assess the feasibility of using mixing rules to model bonds at solid interfaces, particularly in the context of interfacial heat transport.

While interatomic potentials with mixed parameters may be able to capture the overall energetics of solutions and alloys, effectively averaging over the interactions across the system, it's less clear that they can capture the specific energetics of discrete heteropolar bonds that would exist at an interface. There is some evidence that this may be feasible, given some partial success capturing the energetics of some point defects in SiC [93,98] or replicating the interfacial energy of BN-C interfaces [99]. However, in the context of thermal transport at an interface, the ultimate arbiter of success is how well a potential can capture the underlying IFCs in the interface region. Indeed, the use of mixing rules with Tersoff has failed to replicate the elastic constants of bulk compounds [93,100], suggesting that replicating interface IFC2s may be particularly challenging. Here we use the previously discussed *ab initio* interface IFC2s for the Ge/GaAs interface to assess how likely such an approach is to succeed.

As mentioned in Section I, empirical potentials often struggle to replicate *ab initio* thermal properties and underlying IFC2s in bulk systems, let alone in interface systems. Rather than directly comparing IFC2s produced by a mixed potential to corresponding *ab initio* values, we can generalize our results by recasting the problem as an attempt to replicate the relationship between bulk IFC2s and their corresponding interfacial IFC2s. In this sense, we are considering the mixing of resulting interface IFC2s rather than the mixing of potential parameters, directly examining how interface properties can be reconstructed from bulk properties. As a limiting case, *ab initio* IFC2s could be used to establish an exact relationship between bulk (Ge, s-GaAs) and interface (Ge/GaAs) IFC2s, defining a unique weight for each interfacial IFC2 that relates its value to corresponding bulk values:

$$\Phi_{i,\alpha,j,\beta}^{Ge/sGaAs} = \Phi_{i,\alpha,j,\beta}^{Ge} + w_{mix}\left(\Phi_{i,\alpha,j,\beta}^{sGaAs} - \Phi_{i,\alpha,j,\beta}^{Ge}\right), \quad (1)$$

where $w_{mix}$ is allowed to take on any value and $\Phi_{i,\alpha,j,\beta}$ is an arbitrary IFC2 indexed by atoms $i$, $j$ and cartesian components $\alpha,\beta$. Ideally a mixed potential would be able to approximate this relationship, and so long as the bulk IFC2s are not too far from their *ab initio* values, this would provide reasonable confidence in the accuracy of the mixing rules approach.

In general, mixed potentials can produce arbitrarily complex relationships between bulk and interfacial IFC2s as a consequence of the nonlinearity of many classical potentials, and can thus conceivably approximate the exact *ab initio* relationship. However, since mixing rules are typically applied in circumstances where the exact *ab initio* relationship is unknown, there would be little *a priori* information that can be used to validate a mixed potential or inform the development of better mixing rules. Reflecting this lack of knowledge, we instead consider the possibility that a simple relationship exists between bulk and interface *ab initio* IFC2s in hopes that a simple heuristic may be used to validate the performance of mixed potentials.

Specifically, we study the performance of a homogenous mixing scheme that predicts interface IFC2s as the arithmetic or harmonic mean of the corresponding bulk values. A number of past studies have adopted this approach to model interfacial IFCs in simple Green's function calculations of thermal transport [101–103]. However, there is already prior evidence that the use of the arithmetic or harmonic mean to model interfacial IFC2s does not provide good predictions of thermal conductance [74]. Thus we also consider a more general weighted mean, rearranging Eq. 1 into a more common form,

$$\widehat{\Phi}_{i,\alpha,j,\beta}^{Ge/GaAs} = w_{mix}\Phi_{i,\alpha,j,\beta}^{sGaAs} + (1 - w_{mix})\Phi_{i,\alpha,j,\beta}^{Ge}, \quad (2)$$

and identifying a single optimal weight $w_{mix} \in [0,1]$ to use for all interfacial IFC2s that minimizes the mean absolute percent error (MAPE) of predicted IFC2s relative to their *ab initio* values.

Note that our use of a simple homogenous mixing scheme can be considered a special case where mixing rules are applied to bulk TEPs, such that the mixing relationship between potential parameters is exactly equivalent to the mixing relationship between bulk and interfacial IFC2s.

Table 1. Mean absolute percent errors of predicted interface IFC2s using different means of respective bulk IFC2 values, with and without outliers.

| Outliers | Mean Type | $w_{mix}$ | Mean Absolute Percent Error of Predicted IFC2s | | | |
|---|---|---|---|---|---|---|
| | | | 1NN | 2NN | 3NN | All |
| **All Data** | Harmonic | - | 19% | 1390% | 170% | 621% |
| | Arithmetic | 0.5 | 20% | 286% | 155% | 191% |
| | Optimal Weight | 0.4 | 21% | 218% | 151% | 163% |
| **Outliers Removed** | Harmonic | - | 19% | 160% | 102% | 114% |
| | Arithmetic | 0.5 | 20% | 112% | 71% | 81% |
| | Optimal Weight | 0.5 | 20% | 112% | 71% | 81% |

The results of these homogenous mixing approaches are summarized in Table 1. As discussed in the Section SIII of the SI, outlier values corresponding to very small reference IFC2s can distort the MAPE, and so we also present results with outliers excluded. Overall, the MAPE is fairly high, demonstrating that a simple homogenous mixing scheme is unable to accurately replicate interfacial IFC2s from bulk values, and that more detailed schemes would be necessary to describe the *ab initio* relationship.

To better understand this result and generalize its implication, we focus on the exact relationship between specific *ab initio* interface IFC2 tensors and their corresponding bulk values. For each unique IFC2 element, we compute its normalized value according to a slightly re-arranged version of equation 1:

$$w_{mix} = \widetilde{\Phi}_{i,\alpha,j,\beta}^{Ge/GaAs} = \frac{\Phi_{i,\alpha,j,\beta}^{Ge/GaAs} - \Phi_{i,\alpha,j,\beta}^{Ge}}{\Phi_{i,\alpha,j,\beta}^{sGaAs} - \Phi_{i,\alpha,j,\beta}^{Ge}}. \tag{3}$$

With this normalization, a value of zero would correspond to the bulk Ge value, a value of 1 would correspond to the bulk s-GaAs value, and 0.5 would be exactly in between the two values (i.e., the arithmetic mean).

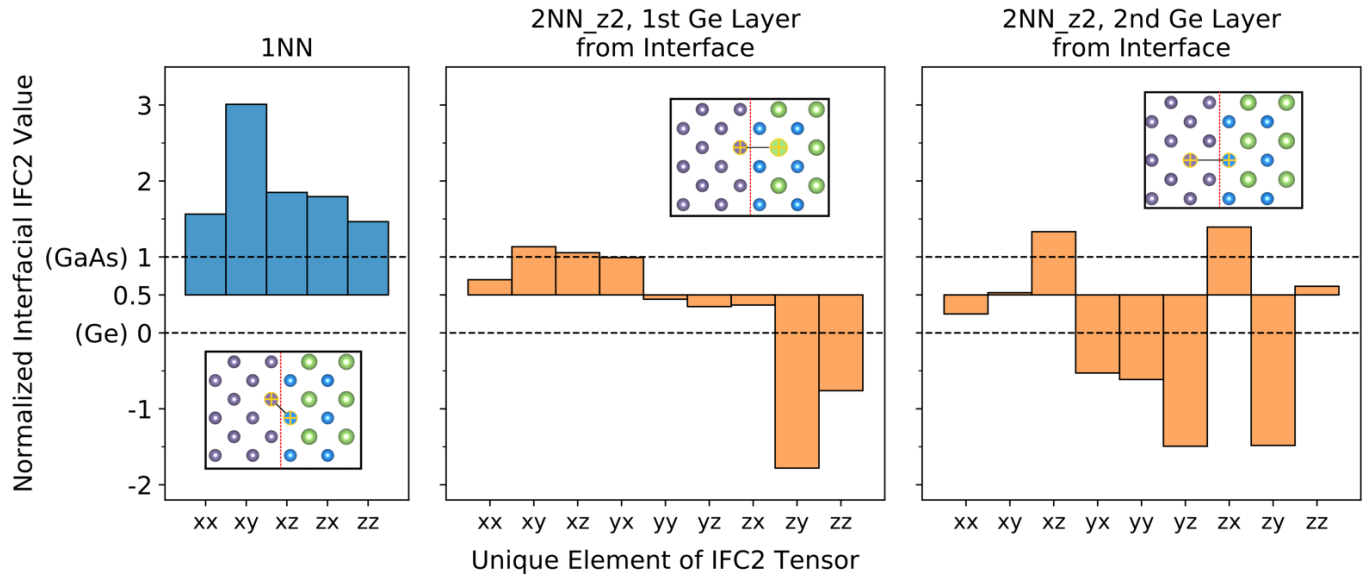

*Figure 2. Normalized values of individual IFC2s for three types of interfacial IFC2 tensors. Note that the symmetry of the 1NN IFC tensor reduces the number of unique IFC2 elements from 9 to 5. The insets provide representative examples of atomic pairs associated with each tensor.*

The results of this normalization are presented for 1NN and 2NN IFC2s in Figure 2, and a corresponding plot for 3NN in Fig. S7 of the SI. A rather nontrivial relationship between bulk interface IFC2s is revealed from these plots. Notably, the majority of IFC2s are not bounded by either the bulk Ge or bulk s-GaAs IFC2 values. This helps to explain the poor performance of the homogenous mixing scheme using mean values of the bulks, as such an approach can only predict values bounded by respective bulk values (i.e., between 0 and 1 in Figure 2.)

Interestingly, all 1NN interface IFC2s adopt values greater than 1, indicating their preference for IFC2 values that are more GaAs-like (and in fact significantly surpass them), something that was not *a priori* obvious. The 2NN IFC2 tensors exhibit a complicated mixture of more Ge-like IFC2s and more GaAs-like IFC2s. When comparing the 2NN IFC2 tensor whose associated atomic pairs are more on the Ge side (right plot) than on the GaAs side (middle plot), most IFC2 elements trend more Ge-like as one would expect, but a few elements (xz, zx, and zz) become more GaAs like. Additionally, SI Fig. S8 further highlights the complexity of the relationship between bulk and interfacial IFC2s, demonstrating that interfacial IFCs2 can sometimes be of a different order of magnitude and/or sign as their respective bulk values.

All of this demonstrates that, even for a relatively simple and high symmetry system like Ge/GaAs, a rather complex relationship can exist between *ab initio* bulk and interfacial IFC2 values. One consequence of this is the inability to use simple heuristics to evaluate the performance of mixing rules in arbitrary potentials. For example, if most *ab initio* interface values had been bounded by their bulk values, we could leverage this information to check whether a mixed potential produced interface IFC2s that were themselves bounded by the bulk values of the unmixed potentials. Such a heuristic would have been a convenient check that could help constrain the IFC errors of mixed potentials, but is unfortunately not justified.

It is still conceivable that a mixing rule scheme could reproduce a similar bulk-interfacial IFC2 relationship as exhibited here, or at least capture the leading order features of the relationship. Indeed, one could use the presented results for Ge/GaAs and choose to use GaAs potential parameters to model Ge-As interfacial bonds, assuming that the thermal transport properties will be determined primarily by the largest harmonic IFCs. However this kind of insight is materials system-dependent, and as emphasized, mixing rules are typically applied in situations where such *ab initio* data doesn't exist. At a minimum, a more comprehensive study would be needed investigating the relationship of *ab initio* bulk and interfacial IFC2s for a variety of material systems, to attempt to obtain more sophisticated heuristics to guide the validation and development of arbitrary mixed potentials. Otherwise, one has to rely on the physical justification of existing mixing rules, which is generally rather weak, and trust that it can reproduce a complex relationship between bulk and interface IFCs.

A far more principled approach is to explicitly treat interface interactions by fitting potentials to *ab initio* interface data, as in the case for MLIPs or TEPs. While it may still be difficult for a given MLIP to accurately reproduce IFC2s, especially smaller ones, there is a clear path of improvement by fitting to more/better training data and incorporating more sophisticated ML models. Given this and other limits of mixing rules (e.g., the need to use the same functional form to model both constituent materials, and the fact that even bulk thermal properties are often poorly reproduced by analytic potentials), the use of analytic potentials with applied mixing rules to study thermal transport in interface systems is poorly justified.

# IV. Explicitly Modeling Interface Interactions

## A. Interatomic Potential Development and Model Selection

As computational capabilities have grown, it is now possible to generate reasonable, if not large, amounts of *ab initio* data for bulk and interface structures. This data can be used directly in the development of interatomic potentials for use in simulating thermal transport at interfaces. While such potentials represent a significant improvement over classical potentials using mixing rules, they still require judicious evaluation in order to ensure that bulk and interfacial *ab initio* thermal properties are sufficiently replicated by the potential. Moreover, the large flexibility provided by MLIPs and TEPs, which enable them to approximate the *ab initio* PES, unfortunately increases the likelihood of dynamic instabilities, especially when modelling large systems for long time scales.

In this section we investigate two different potentials that can be used to explicitly treat interface interactions. First, we consider a simple representative MLIP known as the spectral neighbor analysis potential (SNAP) [69], a linear model of bispectrum component descriptors fit using ordinary least squares. Second, we consider a potential based on the Taylor expansion of the potential energy surface. Such TE-style potentials are a tantalizing prospect for thermal transport MD studies because they are directly parameterized by *ab initio* IFCs, allowing for exact replication of those force constants by definition. The key challenge in deploying TE-style potentials to model interface systems is the need to stitch together different IFC sets describing the two bulk regions and the interface region, as discussed in Section II, while avoiding dynamic instabilities and adequately modeling anharmonicity.

Our solution to this challenge is as follows. When modelling a large interface system using MD, the bulk Ge, bulk s-GaAs and interface regions are primarily described by their respective IFC2 sets. For atom pairs that straddle the boundary between interface and bulk regions, we simply choose to assign the appropriate bulk IFC2 value, as justified by the demonstrated convergence of IFC2s with the length of the DFT interface system in Section II. However, this would still induce deviations to the acoustic sum rule (ASR) for the hessian of the combined large system and possibly result in instability if a naïve TEP formulation was used. Instead, we use the translationally-invariant TEP (TITEP) developed by Rohskopf et al. [51] which automatically satisfies the ASR by its functional form.

Presently, the TITEP formalism can only be used with harmonic force constants, and a separate potential is needed to model the anharmonic portion of the force. Leveraging the aforementioned flexibility of MLIPs, we choose to use a single SNAP to model the anharmonic portion of both bulk and interface regions. To obtain this potential, a SNAP is fit to the "anharmonic portion" of *ab initio* forces from bulk and interface structures, where the anharmonic forces are defined as the difference between the full DFT force vector and that predicted by the harmonic TITEP, i.e., $F_{anharm} = F_{DFT} - F_{TITEP}$.

The existence of this additional anharmonic potential unfortunately introduces a small perturbation in the IFC2s, resulting in a combined potential whose harmonic force constants no longer exactly replicate their *ab initio* values. However, it is possible to correct for this perturbation and recover nearly exact IFC2s, as had been done in the prior work by Rohskopf *et*

*Table 2. Mean percent error of predicted force vectors and mean absolute error of force components in parentheses, for chosen potentials evaluated against the test set.*

|  | **Bulk Ge** | **Bulk s-GaAs** | **Ge-GaAs Interface** | **All** |
|---|---|---|---|---|
| **Pure SNAP** | 7.51%<br>(38 meV/Å) | 9.54%<br>(46 meV/Å) | 11.48%<br>(53 meV/Å) | 9.51%<br>(46 meV/Å) |
| **Anharmonic SNAP + TITEP** | 7.09%<br>(38 meV/Å) | 8.28%<br>(42 meV/Å) | 13.72%<br>(65 meV/Å) | 9.68%<br>(48 meV/Å) |

*al.* for bulk systems [51]. In fact, this procedure can also be applied as a *post-hoc* correction to any arbitrary potential, including pure SNAP fit to total forces. Thus, we consider the application of this correction procedure both to pure SNAP and to our anharmonic SNAP + TITEP potentials. The details of this correction procedure are provided in the SI.

Using the previously described bulk Ge, bulk s-GaAs, and Ge/GaAs interface structures, fitting data composed of ~1930 unique bulk Ge configurations, ~1970 bulk s-GaAs configurations, and ~3600 interface configurations was generated using random displacements, *ab initio* molecular dynamics (AIMD), and by sampling TITEP trajectories. The details of this fitting data and its generation are provided in the SI.

For both the pure SNAP (p-SNAP) and the anharmonic SNAP + TITEP (aS+T), the cutoff and number of parameters of the SNAP were varied to evaluate the best combination. Model selection was determined based on stability criteria and force performance on a holdout validation set, the details of which are provided in the SI. For p-SNAP, the chosen potential used a 5.133 Å cutoff and the highest number of parameters tested in this study, i.e., $J_{max}=4$ according to the terminology of reference [104]. Unfortunately, all but one aS+T potentials that were initially tested produced instabilities in the interface system. After some additional *ad-hoc* testing, it was determined that stable aS+T potentials could be produced if the anharmonic SNAP was only fit to random displacement data, without significantly compromising the force error. (Note that this was not due merely to the reduction in number of fitting data, as that was also tested while keeping the proportional composition of the training data the same; the potentials remained unstable.)

When the IFC2 correction procedure was applied to both p-SNAP and aS+T, nearly all potentials remained stable for bulk systems. However, nearly all corrected potentials were unstable when modeling the interface system, though an earlier preliminary potential, described in Section SVI of the SI, was able to preserve stability for both bulk and interface systems. Thus, while it appears that the correction procedure is feasible, it is a rather fragile approach. As such, we chose instead to use an uncorrected aS+T, which also happens to have a better overall force error than the only stable corrected potential tested in this work. The associated SNAP is fit only to random displacements, with a cutoff of 5.133 Å and $J_{max}=4$.

The resulting force performance of each chosen potential against the test set is described in Table 1. Forces are evaluated by the mean percent error of their predicted force vectors, defined in SI equation S2, and mean absolute error of force components. Overall, the two approaches produce roughly similar force errors, with less than 10% mean percent error in force vectors for bulk systems for both potentials, and 11.5% and 13.7% force errors for the interface produced by p-SNAP and aS+T, respectively.

## B. Interatomic Potential Evaluation of Thermal Properties and IFCs

A common and useful way to evaluate the bulk harmonic properties of an interatomic potential is to compute its predicted phonon dispersion. Figure 3(a) presents the predicted phonon dispersions for both p-SNAP and aS+T alongside the DFT reference. While both potentials demonstrate good fidelity with respect to the DFT dispersion, the p-SNAP solution has more appreciable deviations than aS+T, especially for the higher-frequency optical branches.

Quantifying how well anharmonic effects are captured by a given potential is important given their role in determining thermal conductivity. This is especially true for the anharmonic SNAP + TITEP solution, as the fitted SNAP portion of the potential is meant to capture the anharmonicity directly. A simple way of evaluating potentials in this context is to directly compute thermal conductivity using the Boltzmann transport equation (BTE) under the relaxation time approximation (RTA), which takes as input the predicted second and third order IFCs (IFC3s) of the respective bulk material. Figure 3(b) present the computed thermal conductivities as a function of temperature for both potentials and for both bulk systems, compared to *ab initio* results. The aS+T potential exhibits excellent agreement with the reference DFT solution, with an 11.3% and 2.0% error for bulk Ge and bulk s-GaAs respectively. In contrast, p-SNAP is noticeably more deviant, with errors of 33.7% and 20.7% for bulk Ge and bulk s-GaAs respectively.

Thus, with respect to bulk thermal properties, aS+T generally outperforms the p-SNAP implementation, at least with regards to predicted thermal properties like dispersion and thermal conductivity. This is likely due to the direct incorporation of IFC2s within the potential, in spite of the deviation the anharmonic SNAP introduces in the overall harmonic IFCs for the aS+T solution. Interestingly, the excellent performance of aS+T with respect to computed thermal conductivity suggests the anharmonic SNAP approach also does a better job at capturing IFC3s than pure SNAP (though this is not immediately evident when directly evaluating IFC3 predictions, see Fig. S13 and Fig. S14 of the SI).

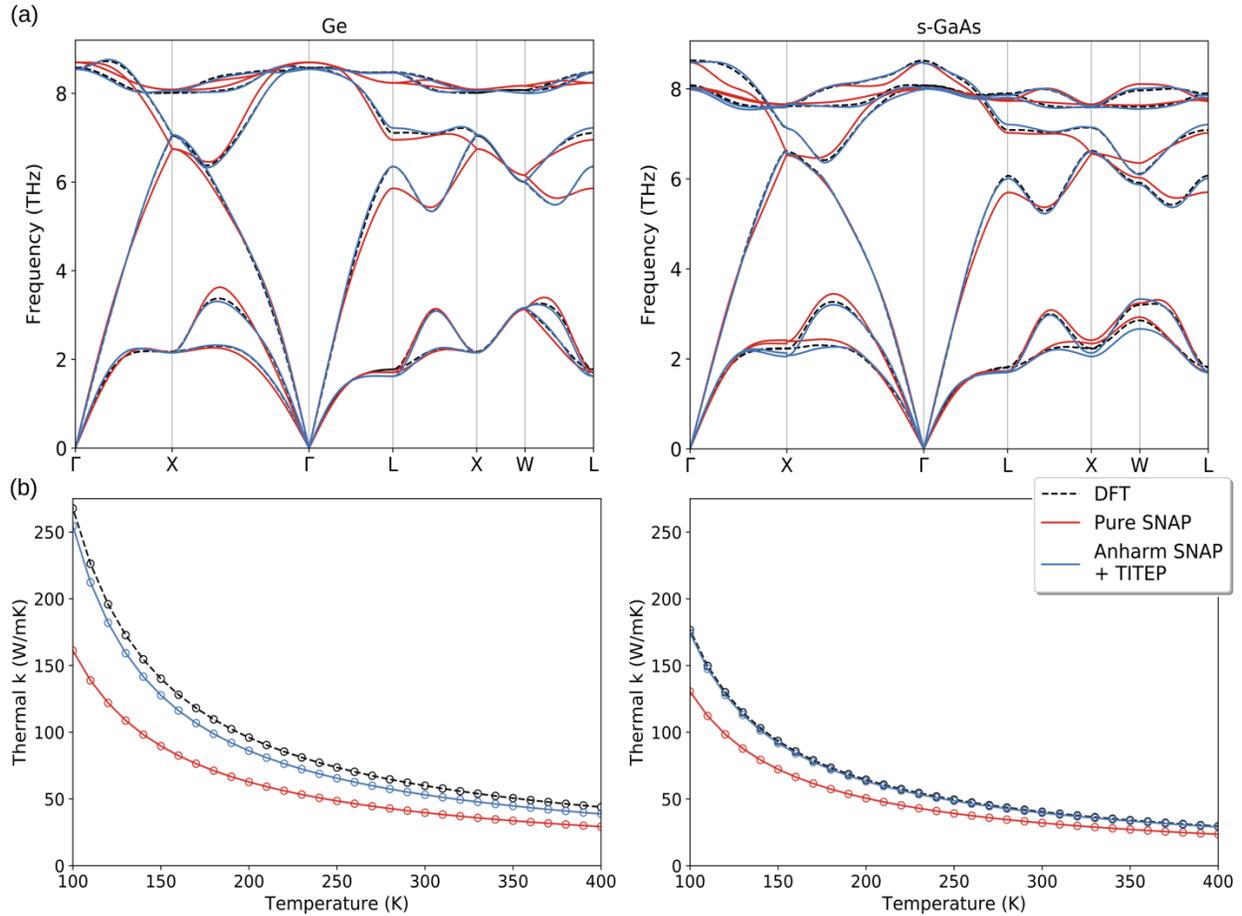

*Figure 3. (a) Predicted and reference DFT phonon dispersion plots. For s-GaAs, IFCs were obtained from the tetragonally distorted structure, but are mapped back to the undistorted but strained s-GaAs structure for ease of visualization (b) Predicted and reference DFT thermal conductivity computed with BTE and the RTA*

Next, we turn to evaluating the thermal performance of these potentials within the interface region. While it is technically possible to produce a dispersion for an interface superlattice system, such plots often consist of numerous and relatively flat bands due to the broken symmetry of the interface region (e.g., [105]). Instead, we choose here to directly compare the predicted IFC2s themselves, exploring a number of different visualization and comparison strategies to help inform our evaluation.

To start, we plot predicted vs. reference IFC2s using a symmetrical log-log plot in Figure 4, which allows us to include sign information for data that spans multiple orders of magnitude. A number of important features stand out. As expected, IFC2s with larger reference magnitudes (typically 1NN) are generally much more accurately captured than those with smaller magnitudes (i.e., 2NN and 3NN terms). Notably for both potentials, a number of the smaller predicted IFC2s have flipped signs relative to the reference DFT IFC2s, though it's unclear what, if any, implication this may have. From these plots, at a qualitative level, aT+S appears to provide a better fit than p-SNAP, but it is challenging to meaningfully compare these potentials using such an approach.

In order to quantitatively compare the performance of p-SNAP and aT+S in terms of the raw IFC2 values in the interface region, we use a box plot to describe the absolute percent errors of each potential binned by their nearest neighbor classification. The resulting plot, seen in Figure 5, demonstrates that aS+T clearly performs better than p-SNAP, though overall median errors are similar. Note that for 1NN, *median* absolute percent errors (APE) are below 2% for each potential, while for 2NN median APEs are ~11% and ~7% for p-SNAP and aS+T respectively. Nonetheless, there are numerous 2NN IFCs whose APE exceeds 100% for both potentials and 3NN IFC2s exhibit much higher errors.

While these results are impressive for the aS+T potential, our solution unfortunately suffers from a critical drawback, namely, it is not stable when simulating a large interface system with bulk and interface regions. As discussed, it is in principle possible to leverage the strong IFC2 convergence of the interface system to stitch together the bulk and interface hessians, using the bulk IFC2s for bonds in between the bulk and interface region. We had previously demonstrated the viability of this potential with an earlier version of this potential, developed using a slightly different Ge/GaAs interface and a smaller fitting set generated with different DFT data (using the generalized gradient approximation). This potential is described in SI section SVI.

However, when the same approach was adopted with this potential, the resulting combined potential was unstable. Interestingly, it appears that the source of the instability is not with the stitching procedure. Rather, the TITEP (and TEP) that models the interface, while stable when modeling the original DFT-sized interface system, is unstable when a supercell of this DFT system is created in the cross-plane directions. It is unclear exactly what causes this, as the same issue did not arise in the earlier potential. Nonetheless, it once again demonstrates the fragility of this approach, which should be taken into account in future studies using these methods.

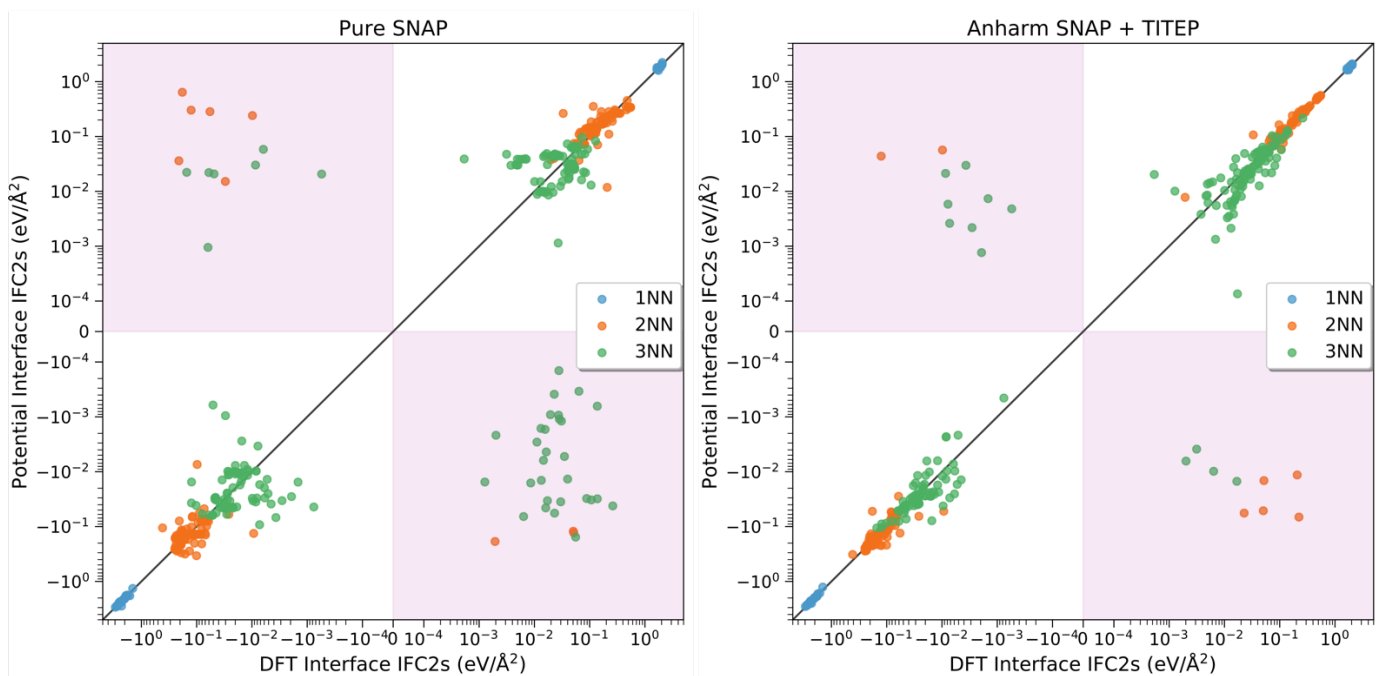

*Figure 4. Symmetric log-log plot of predicted harmonic IFCs against DFT for the entire interface system. The red shaded region indicates regions of sign flipping.*

In contrast, the p-SNAP potential is stable when used in large systems. Although it exhibited somewhat worse thermal performance compared to aS+T, it should be emphasized that these results are still quite good, especially in regards to large IFCs. Moreover, plenty of additional modifications to this potential could be pursued, including a more complete exploration of its hyperpameters, the choice of fitting data, and use of regularization in the fitting procedure.

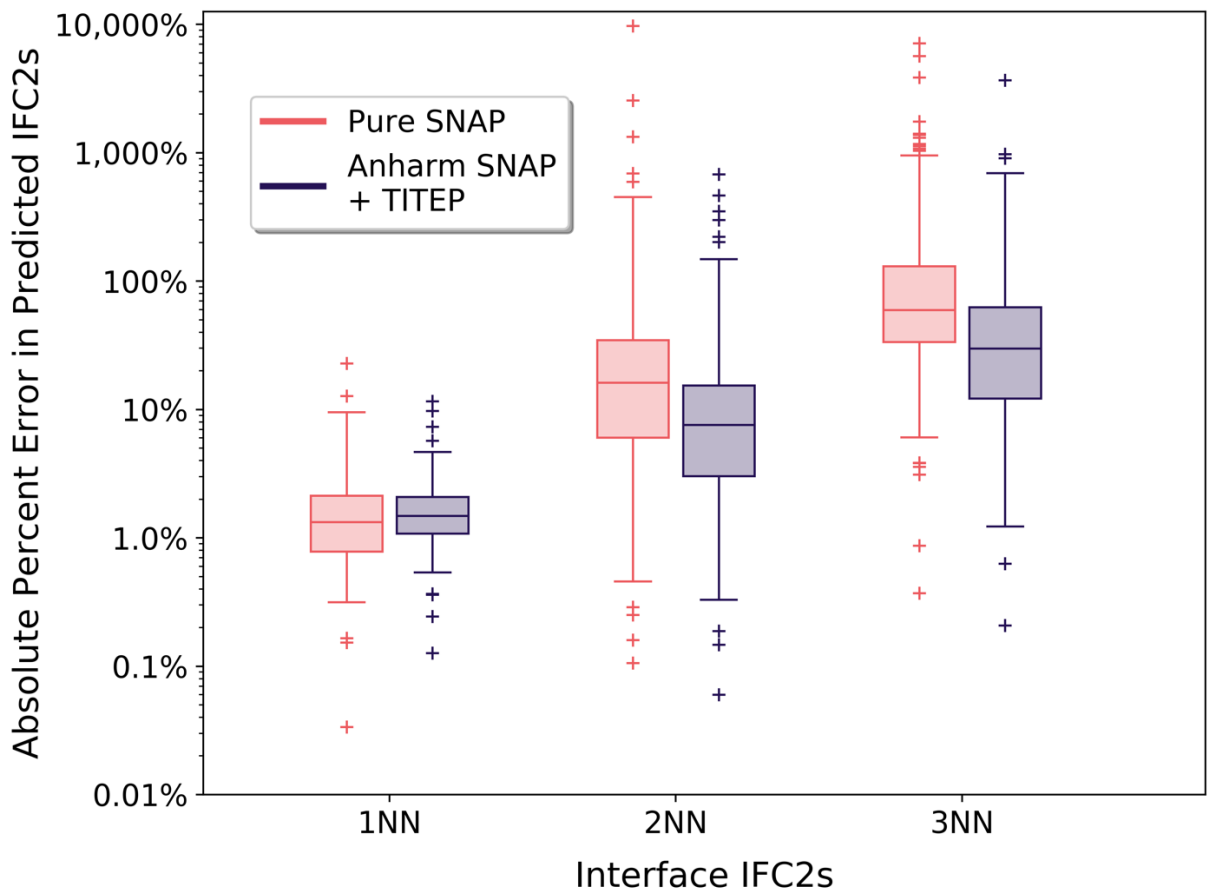

*Figure 5. Box plot of the absolute percent error of predicted interface IFC2s for p-SNAP and aS+T, binned by nearest neighbor. Data is log-transformed before determining the respective median and quantiles. Box plot whiskers correspond to Turkey's original definition [106].*

## IV. Discussion and Conclusions

The central focus of this work is on the development of interatomic potentials to model thermal transport for interface systems. Three main facets of this problem were explored: the ability to generate reliable *ab initio* reference data with converged IFCs, the general necessity of explicitly treating interface interactions, and the ability to meaningfully evaluate the potentials with respect to their bulk and interfacial thermal properties.

In regards to the convergence of interface IFC2s towards their bulk values, it is important to emphasize that our results are unique to the Ge/GaAs system, which has high symmetry and exhibited minimal distortions. Other lower symmetry systems that may also include significant distortions are unlikely to exhibit similarly quick convergence. Indeed, for such systems it may be difficult to even obtain a reliable IFC2 set in the first place, as the low symmetry can lead to a dramatic increase in the number of IFC2 parameters. While this does not prohibit the generation of *ab initio* fitting data and the development of a MLIP, future researchers should recognize the uncertainty this can cause in the reliability of the produced potential. Ultimately, such an analysis should be attempted when possible, as it provides a robust measure of how well a DFT interface model represents the ideal semi-infinite case.

Given the focus on IFCs in this paper, it is important to highlight the careful attention needed in how they are obtained. As derivatives of forces, IFCs can be particularly sensitive to the underlying electronic parameters, especially the density of the k-point sampling. This is especially important for smaller IFCs associated with longer range and higher order terms. It may be valuable to treat small IFCs as data with some intrinsic uncertainty, which could affect our analysis of IFC convergence and the thermal property evaluation of interatomic potentials.

In assessing the viability of mixing rules to produce reasonable interface IFCs, we demonstrated the imposing challenge such an approach faces in attempting to meaningfully replicate the relationship of bulk to interfacial IFC2s that is exhibited in *ab initio* calculations. While we cannot conclude it is impossible, there is little *a priori* information that could be used to inform the choice of a mixing scheme or their evaluation. Moreover, since most mixing rules schemes used to model realistic interfaces are rather simple and have rather weak physical justifications, it seems particularly unlikely that such methods will provide reasonable IFCs. Our core argument is that, given these weaknesses, researchers should carefully consider whether using mixing rules is meaningfully justified for their circumstances, and if not, that the time and resources are likely best spent developing MLIPs or related potentials to model interfacial heat transport, rather than attempting to validate and optimize new mixing rule schemes.

In this regard, we presented results for two different kinds of potentials that are fit to *ab initio* data of Ge/GaAs bulk and interface systems: a pure SNAP and a hybrid anharmonic SNAP overlayed on a TITEP. The resulting potentials are able to reproduce bulk thermal properties and interface IFC2s reasonably well, though the anharmonic SNAP + TITEP outperformed pure SNAP in this regard. However, we were unable to reliably obtain a stable anharmonic SNAP + TITEP that could simulate a large interface system, limiting its practical use in thermal MD studies. Additionally, for the aS+T solution, it appears as though fitting to random displacement data is more likely to produce a stable potential for individual bulk and interface systems with optimal force error, a behavior that deserves further exploration in later studies.

Ultimately, MLIPs are particularly promising for the study of interfacial heat transport, given their relative ease-of-use and overall accuracy, so long as sufficient fitting data are generated. It's important to note that SNAP is one of many possible MLIPs that can be used, and appreciable refinement is likely possible with better models. In contrast, the proposed TITEP + anharmonic MLIP is, as discussed, a somewhat fragile solution that needs further study before being more widely used. Nonetheless, its superior thermal performance should provide motivation to continue developing TE-style methods, with a focus on augmentations that can improve stability when running MD. Indeed, prior work suggests that higher order TEPs with short anharmonic cutoffs are a feasible and perhaps more viable route to develop TE-style potentials for interfacial heat transport [67].

Further study will be needed to evaluate the performance of these and other potentials on key properties related to interfacial heat transport, namely thermal boundary conductance and its modal decomposition. Likewise, the anharmonic properties of interfacial heat transport were not directly studied here as we were unable to obtain reliable third order IFCs for the interface system. It may also be valuable to establish a better correspondence between the performance of a potential directly with respect to IFCs and with respect to downstream thermal properties of interest. Such a correspondence could be informed by systematic sensitivity studies, relating changes to underlying force constants to properties like thermal conductivity and TBC.

## Acknowledgements

The authors acknowledge support from the United States Office of Naval Research Multidisciplinary University Research Initiative Grant Number N00014-18- 1-2429, as well as the National Science Foundation for Career Award Grant Number 1554050.

# Development of Interatomic Potentials to Model the Interfacial Heat Transport of Ge/GaAs
## *Supplementary Information*

Spencer Wyant[1], Andrew Rohskopf[2], Asegun Henry[2]

[1] Department of Materials Science and Engineering, Massachusetts Institute of Technology, 77 Massachusetts Avenue, Cambridge, Massachusetts 02139, USA

[2] Department of Mechanical Engineering, Massachusetts Institute of Technology, Cambridge, Massachusetts 02139, USA

## Section SI: Methods
### A. Details of DFT Calculations

All *ab initio* data was generated using density functional theory (DFT) with the plane-wave based Vienna ab initio Simulation Program (VASP) [1–3], using projector augmented wave pseudopotentials ($4s^24p^2$, $4s^24p^1$, $4s^24p^3$ valence electrons for Ge, Ga, As). The exchange and correlation portion of the DFT functional was treated under the local density approximation (LDA) as parameterized by Perdew and Zunger [4]. Unless otherwise stated, all DFT calculations were computed with strict electronic parameters to ensure forces were well converged, with a 600 eV plane wave cutoff and an electronic convergence of 1e-8 eV. Gaussian smearing was used to handle partial state occupancy, with a smearing parameter of 0.01 eV.

The Brillouin zone was sampled using a k-point mesh generated by a gamma-centered Monkhorst-Pack scheme. The grid sizes were chosen to produce similar reciprocal spacings for the bulk and interface models. For DFT calculations used to generate force and energy fitting/testing data, a k-point mesh of 2x2x2 and 3x3x1 was used to model bulk and interface models, respectively. Since IFCs can be more sensitive to the underlying electronic parameters (as they are derivatives of energetic properties), a denser k-point mesh was used for finite displacement calculations: 4x4x4 for bulk systems and 4x4x1 for interface systems. The denser k-point meshes were also used in the initial relaxation of the bulk and interface systems.

All relaxations use the conjugate gradient algorithm with a force convergence of 1e-5 eV/Å. Initially, bulk Ge and bulk GaAs were relaxed in their ground state diamond-cubic and zincblende structures, respectively, obtaining lattice parameter values of 5.647Å for bulk Ge (5.652 expt. [5]) and 5.627Å (5.648 expt. [5]) for bulk GaAs. To obtain s-GaAs, the in-plane lattice parameters are fixed to the bulk Ge values, and the out-of-plane lattice parameter was manually relaxed, inducing a small tetragonal distortion of -0.0382 Å relative to its initial relaxed bulk value. Both bulk Ge and bulk s-GaAs use 216 atom 3x3x3 supercells of the conventional cubic unit cell for all fitting data generation and IFC-related calculations.

### B. Training, Validation, and Test Data Generation

*Ab initio* force and energy data, used to fit and evaluate interatomic potentials, were generated in one of three possible ways: random displacements, sampling *ab initio* molecular dynamics (AIMD) trajectories, or sampling TITEP trajectories. For randomly displaced structures, atomic positions are displaced with a magnitude corresponding to a uniform sampling of a range of displacement (0.0-0.06 Å, 0.0-0.1 Å, 0.0-0.2 Å, 0.04-0.1 Å, 0.08-0.14 Å, 0.12-0.18 Å), and in a uniformly sampled random direction (i.e., by sampling azimuthal and polar coordinates according to $\theta = 2\pi u, \phi = \cos^{-1}(2v - 1); u, v \in [0,1)$ )

AIMD data was generated using a two-step process. First, AIMD trajectories were generated primarily under an NVT ensemble, with Born-Oppenheimer dynamics and a Nose-Hoover thermostat, using a timestep of 1 fs. Less stringent electronic parameters were used during AIMD, with only gamma point Brillouin sampling and an electronic convergence of 1e-6 eV. These trajectories were then sampled every 50 fs and the sampled configurations were re-computed with the more stringent electronic parameters previously described, to be consistent with the rest of the data set. Trajectories were typically run for about 8.5-10.5 ps at 300K, 400K, 600K and 900K.

Data generated from sampling TITEP trajectories were produced as follows. TITEPs, parameterized by the respective harmonic IFCs, were separately produced for bulk Ge, bulk s-GaAs, and the relaxed Ge/GaAs interface. NVT MD simulations were run at temperatures ranging from 200K-900K, and snapshots were periodically sampled every 40 ps. These snapshots were then recomputed with DFT to produce *ab initio* force and energy data. Sampling TITEP trajectories is similar to methods that directly sample combinations of normal modes of a system under the canonical ensemble, which approximates the phase space explored in AIMD at significantly reduced computational cost.

### C. Fitting, TITEP Correction Procedure, and Model Selection

Potentials investigated consisted of either pure SNAP fit to total forces and energies or a combined potential of SNAP fit to anharmonic forces and energies overlayed on top of bulk/interface TITEPs, as discussed in Section III of the main text. Both types of SNAP are fit using ordinary least squares. The functional form of SNAP is thoroughly described in prior works [6,7], and the details of TITEP are provided in ref. [8].

As discussed in Section III, TITEP needs to be corrected/adjusted when combined with another potential, in order to recover exact IFC2s. The SNAP and TITEP portions of the combined potentials contribute to the total energy in a linear fashion, and as such derivatives of the energy -- forces and force constants -- can also be decomposed linearly into SNAP and TITEP contributions. When TITEP is parameterized by *ab initio* IFC2s, individual force constants obtained with TITEP should be equivalent to respective DFT force constants, up to numerical error. However, SNAP, whether fit to the total force or the anharmonic portion of the force, produces nonzero IFC2s for force constants whose atomic pairs are within the cutoff of SNAP. While these IFC2s tend to be quite small for SNAP fit to anharmonic forces, they nonetheless perturb the overall IFC2s such that $\Phi_{i,\alpha,j,\beta}^{DFT} \neq \Phi_{i,\alpha,j,\beta}^{TITEP} + \Phi_{i,\alpha,j,\beta}^{SNAP}$.

However, this linearity can be leveraged to generate a modified TITEP whose underlying IFC2s are constructed so as to recover the DFT IFC2s for the overall combined potential. This is accomplished by simply subtracting the full hessian obtained from SNAP alone from that obtained with DFT, or equivalently:

$$\Phi_{i,\alpha,j,\beta}^{TITEP\_CORR} = \Phi_{i,\alpha,j,\beta}^{DFT} - \Phi_{i,\alpha,j,\beta}^{SNAP}, \ \forall \ (i,j,\alpha,\beta) \tag{S1}$$

Subtleties exist as to where the force constants or are evaluated, whether at the minimum equilibrium DFT structure, the minimum of the SNAP, or the minimum of the uncorrected combined potential. However, tests of these different choices resulted in minimal difference in resulting forces, force constants, or stability. Note that when force constants are obtained for evaluation (i.e., in Section IV of the main text), they are evaluated at the minimum of the tested potential.

The total data set consisted of 2407 bulk Ge configurations, 2464 bulk s-GaAs configurations, and 4196 interface configurations. From this, a test set composed exclusively of 973 (bulk and interface) AIMD and sample-TITEP data (10.7% of the total data) was split off to be used in the final evaluation of our created potentials. The choice to use only AIMD and sample-TITEP data for the test set stems from an assumption that such data is more representative of a realistic phase space compared to randomly displaced structures. The remaining data is then split between train and validation sets (in a roughly 93%/7% split), the latter being used to evaluate different SNAP hyperparameters. A full breakdown of the data set is provided in SI Tables S1-S3.

Since both the full training data set and a subset consisting of only random displacement (RD) data were used to fit potentials, totals for both the full set and the randomly displaced set are provided.
Section SIV. discusses the model selection of the pure SNAP and anharmonic SNAP + TITEP potentials.

Table S1. Composition of bulk Ge dataset, providing the number of train/val/test configuration for each category of data.

| Bulk Ge Dataset | | | | | |
|---|---|---|---|---|---|
| **Type** | **Temperature** | **Random Disp. Range** | **Train** | **Validation** | **Test** |
| Random Disp. | - | 0.0-0.1 Å | 139 | 11 | 0 |
| Random Disp. | - | 0.0-0.2 Å | 138 | 12 | 0 |
| Random Disp. | - | 0.0-0.06 Å | 93 | 7 | 0 |
| Random Disp. | - | 0.04-0.1 Å | 92 | 8 | 0 |
| Random Disp. | - | 0.08-0.14 Å | 93 | 7 | 0 |
| Random Disp. | - | 0.12-0.18 Å | 92 | 8 | 0 |
| AIMD | 300K | - | 258 | 20 | 65 |
| AIMD | 600K | - | 252 | 19 | 62 |
| AIMD | 900K | - | 194 | 16 | 51 |
| Sample TITEP | 300K | - | 83 | 6 | 21 |
| Sample TITEP | 400K | - | 2 | 7 | 21 |
| Sample TITEP | 500K | - | 83 | 6 | 21 |
| Sample TITEP | 600K | - | 82 | 7 | 21 |
| Sample TITEP | 700K | - | 83 | 6 | 21 |
| Sample TITEP | 800K | - | 82 | 7 | 21 |
| Sample TITEP | 900K | - | 83 | 6 | 21 |
| ***Bulk Ge Total*** | | | **1929** | **153** | **325** |
| ***Bulk Ge (just RD) Total*** | | | **647** | **-** | **-** |

Table S1. Composition of bulk s-GaAs dataset, providing the number of train/val/test configuration for each category of data.

| Bulk s-GaAs Dataset | | | | | |
|---|---|---|---|---|---|
| **Type** | **Temperature** | **Random Disp. Range** | **Train** | **Validation** | **Test** |
| Random Disp. | - | 0.0-0.1 Å | 139 | 11 | 0 |
| Random Disp. | - | 0.0-0.2 Å | 138 | 12 | 0 |
| Random Disp. | - | 0.0-0.06 Å | 93 | 7 | 0 |
| Random Disp. | - | 0.04-0.1 Å | 92 | 8 | 0 |
| Random Disp. | - | 0.08-0.14 Å | 93 | 7 | 0 |
| Random Disp. | - | 0.12-0.18 Å | 92 | 8 | 0 |
| AIMD | 300K | - | 254 | 19 | 64 |
| AIMD | 600K | - | 246 | 20 | 62 |
| AIMD | 900K | - | 248 | 19 | 62 |
| Sample TITEP | 300K | - | 83 | 6 | 21 |
| Sample TITEP | 400K | - | 82 | 7 | 21 |
| Sample TITEP | 500K | - | 83 | 6 | 21 |
| Sample TITEP | 600K | - | 82 | 7 | 21 |
| Sample TITEP | 700K | - | 83 | 6 | 21 |
| Sample TITEP | 800K | - | 82 | 7 | 21 |
| Sample TITEP | 900K | - | 83 | 6 | 21 |
| ***Bulk s-GaAs Total*** | | | **1973** | **156** | **335** |
| ***Bulk s-GaAs (just RD) Total*** | | | **647** | **-** | **-** |

*Table S3. Composition of Ge-GaAs interface dataset, providing the number of train/val/test configuration for each category of data.*

| Ge-GaAs Interface Dataset | | | | | |
|---|---|---|---|---|---|
| Type | Temperature | Random Disp. Range | Train | Validation | Test |
| Random Disp. | - | 0.0-0.1 Å | 185 | 15 | 0 |
| Random Disp. | - | 0.0-0.2 Å | 185 | 15 | 0 |
| Random Disp. | - | 0.0-0.06 Å | 185 | 15 | 0 |
| Random Disp. | - | 0.04-0.1 Å | 185 | 15 | 0 |
| Random Disp. | - | 0.08-0.14 Å | 185 | 15 | 0 |
| Random Disp. | - | 0.12-0.18 Å | 185 | 15 | 0 |
| AIMD | 300K | - | 444 | 35 | 55 |
| AIMD | 400K | - | 440 | 36 | 55 |
| AIMD | 600K | - | 148 | 12 | 19 |
| AIMD | 900K | - | 292 | 23 | 37 |
| Sample TITEP | 300K | - | 166 | 13 | 21 |
| Sample TITEP | 400K | - | 166 | 13 | 21 |
| Sample TITEP | 500K | - | 165 | 13 | 21 |
| Sample TITEP | 600K | - | 166 | 13 | 21 |
| Sample TITEP | 700K | - | 166 | 13 | 21 |
| Sample TITEP | 800K | - | 165 | 13 | 21 |
| Sample TITEP | 900K | - | 166 | 13 | 21 |
| *Ge-GaAs Interface Total* | | | 3594 | 289 | 313 |
| *Ge-GaAs Interface (just RD) Total* | | | 1110 | - | - |

### D. IFC and Thermal Property Calculations

Interatomic force constants were obtained using the direct displacement/finite displacement method as implemented in Alamode (v1.0.2) [9] . For each system, small finite displacements were induced on individual atoms, and the resulting system of force-displacements equations were solved using ordinary least squares. Only symmetry-inequivalent displacements were used with magnitudes of 0.01 Å for harmonic IFCs and 0.02 Å for third order IFCs. For bulk systems, no cutoff was used for harmonic IFCs and a 2NN cutoff (8.4 bohr, 4.445A) was used for 3rd order IFCs. For the Ge/GaAs interface system, a 3NN cutoff (10.0 bohr, 5.29 A) was used for the harmonic IFCs.

Phonon dispersions and thermal conductivities were also computed using Alamode. For bulk s-GaAs, a non-analytic correction is applied using the method proposed by Parlinski [10]. The dielectric tensor and Born effective charges were obtained from a density functional perturbation theory calculation of the conventional unit cell with a 20x20x20 k-point mesh. Thermal conductivity was calculated using the BTE framework under the relaxation-time approximation with a 20x20x20 q-mesh.

# Section SII: IFC Force Convergence

Considering only IFC2s up to 3NN for the (DFT) interface system, there are five different types of IFC2 tensors that can be distinguished based on their nearest neighbor shell and the number of planes the associated atom pairs span in the z direction. Figure S1 provides examples of atomic pairs associated with these five tensors. To provide a visual representation of the convergence of relevant structural parameters, Figure S2 plots the convergence of the distance between planes along the z axis moving away from the interface, relative to respective bulk values. Noticeably, the GaAs side has a slightly better convergence compared to the Ge side, a pattern that is reflected in the IFC2 convergence plot of Figure 1 of the main text. The IFC convergence of Figure 1 is re-plotted for individual IFC2 tensors with all unique IFC2 elements plotted for each plane in Figure

S3, to provide a qualitative sense of the variance in the underlying data. Figure S4 plots the distribution of the absolute magnitude of interface IFC2s associated with 1NN, 2NN, and 3NN bins, demonstrating the roughly two order of magnitude difference between 1NN and 3NN IFC2s.

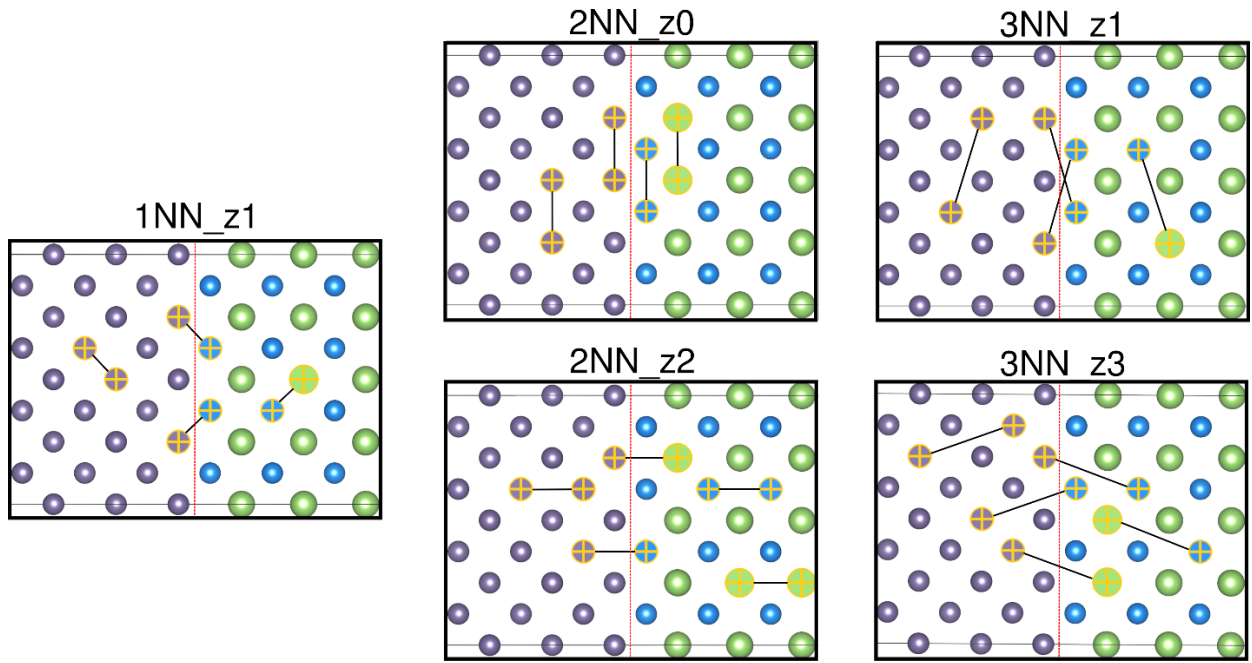

*Figure S2. Representative examples of atom pairs associated with each type of IFC2 tensor. The orientation of the system is the same as Figure 1 of the main text, i.e., these are depictions of the interface in the y-z plane. Note that the displacement vector of most atomic pairs also has a component in the x direction as well, which is not visible due to the projection of these models.*

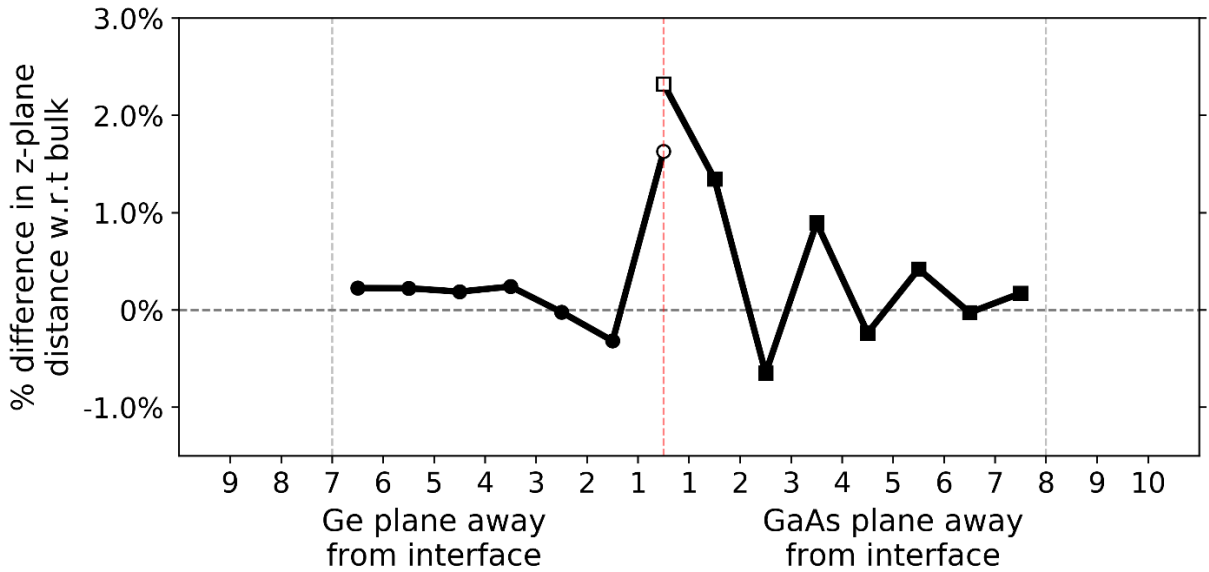

*Figure S2. Plotting the percent difference of distances between atomic planes in the z direction in the interface system relative to respective bulk systems. Circle points represent comparisons to bulk Ge, while square points represent comparisons to bulk s-GaAs. The open markers indicate the z plane distance at the interfacial plane between the Ge and GaAs side of the interface, also indicated by the dashed red line. The gray vertical dashed lines represent midplanes between the interfaces at the middle and at the periodic boundary of the interface system.*

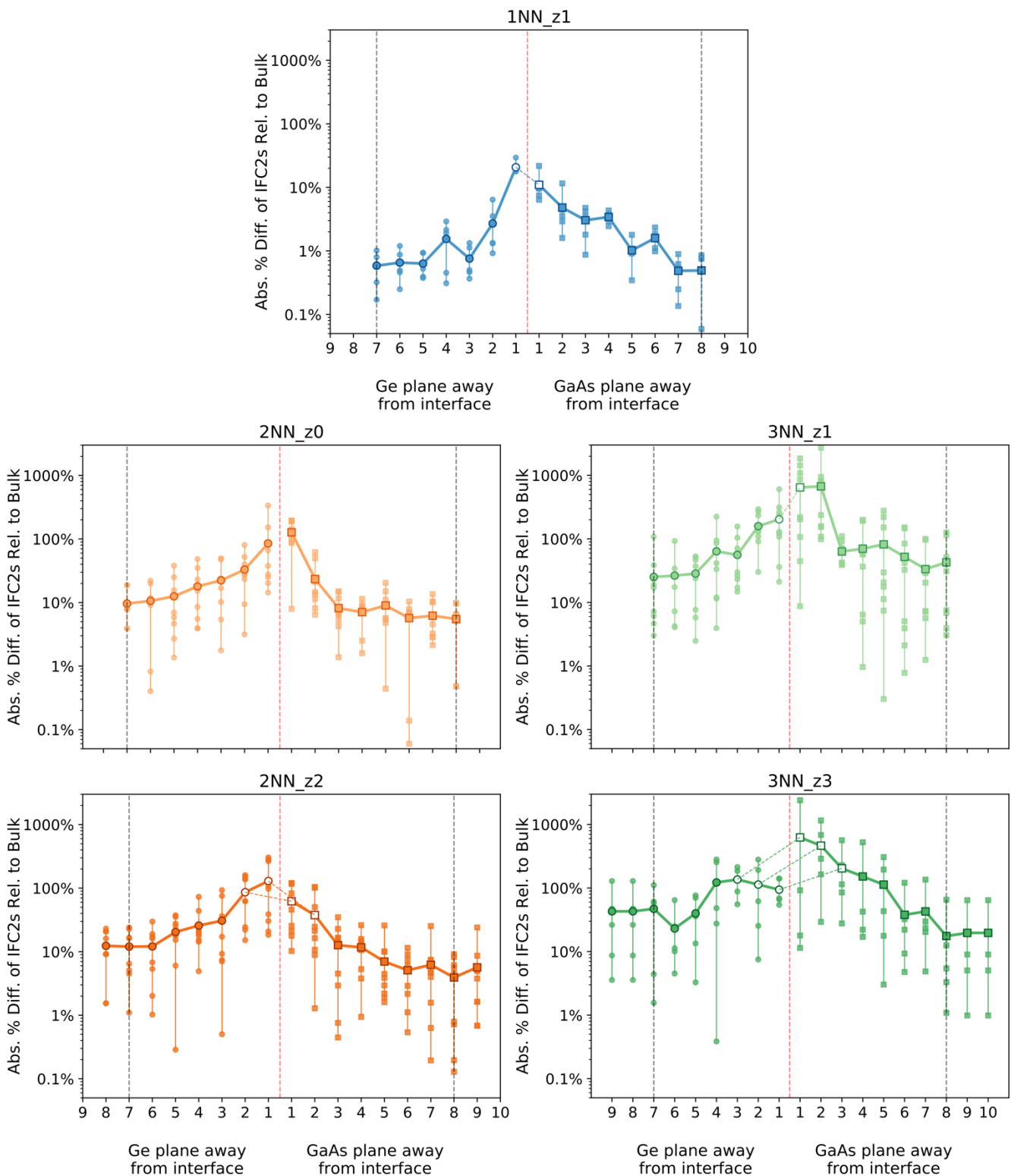

Figure S3. Plots of the absolute percent difference of between IFC2s of the interface system and respective bulk systems, organized by IFC2 type. The same plotting choices described in Fig. 1 of the main text apply to these plots. In addition, mean absolute percent errors (MAPE) are plotted here using markers with darker edge colors. All of the underlying absolute percent errors (APE) that contribute to a given MAPE are plotted at the same x position, with a vertical line connecting minimum and maximum APEs included as visual aid.

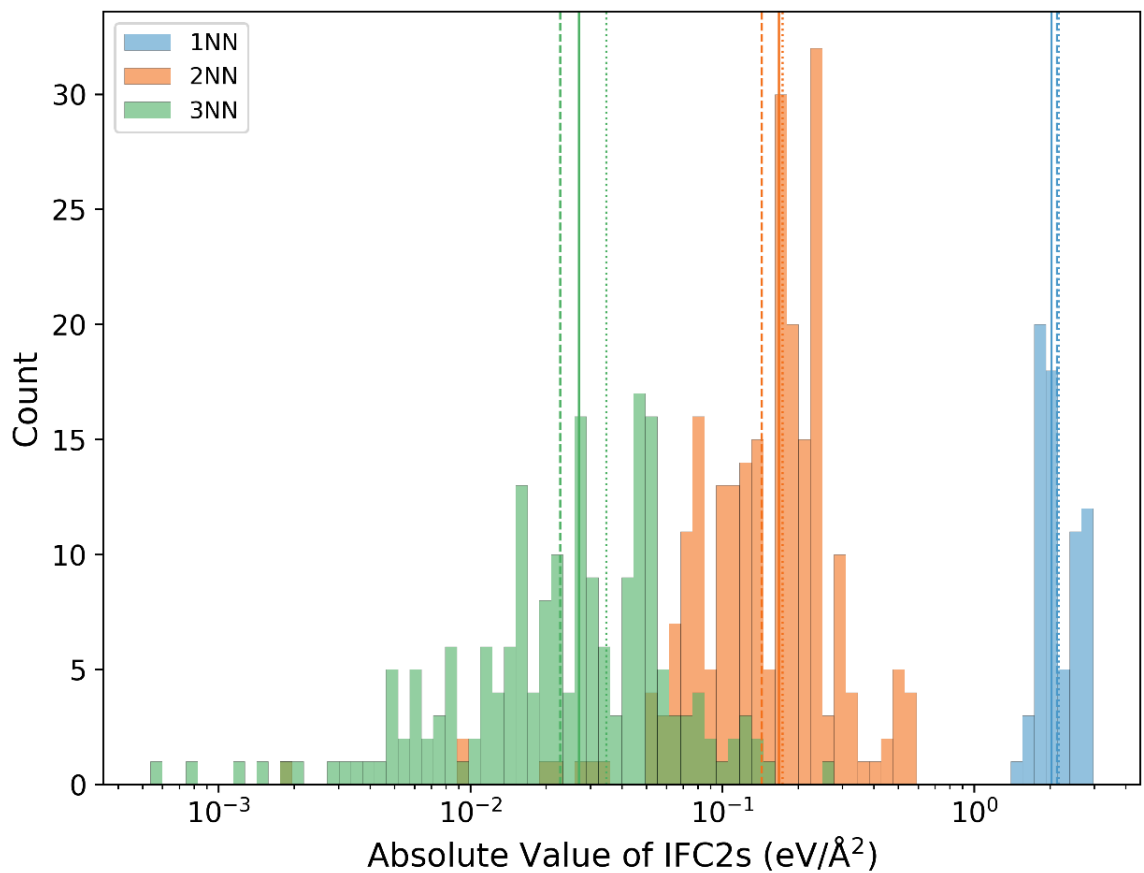

*Figure S4. Histogram of the absolute value of IFC2s for the interface system, demonstrating the order of magnitude difference between IFC2s corresponding to different nearest neighbor shells. The x-axis is plotted on log scale. In addition, the median (solid line), mean (dashed line), and mean of the log-transformed data (dashed line) are plotted for each nearest neighbor bin.*

## Section SIII: Investigating IFC2s at Interfacial Plane

As described in the main text, the relationship between bulk and *interfacial* IFC2s (i.e., those IFC2s whose associated atom pairs span the interfacial plane) is first studied by assessing how well mean combination of *ab initio* bulk values could predict *ab initio* interfacial IFC2s. Specifically, we want to compare the best possible homogenous weighted mean (where homogenous implies that the same weight is used for all interfacial IFC2s). The optimal homogenous weight is chosen to be that which minimized the total mean absolute percent error (MAPE) for all unique 1NN, 2NN, and 3NN interfacial IFC2 elements. The choice to use MAPE rather than mean absolute error (MAE) as the primary error metric is motivated by the desire to assess the performance for all IFC2s, not just the largest 1NN terms, which would otherwise drown out smaller 2NN and 3NN terms if a non-relative error metric is used.

However, the use of absolute percent error (APE) is susceptible to large outlier values when the reference IFC2 terms are very small, and indeed such outlier values do drive up overall MAPE values for our weighted mean predictions. To provide a fairer assessment, we chose to judiciously exclude certain outlier values without shrinking the pool of terms too much. Specifically, interfacial IFC2 elements whose reference *ab initio* values were below 0.01 eV/Å were excluded (see Fig S8 to see which terms), with the exception of one element whose APE values were not particularly high.

To identify the optimal weight, the weight $w_{mix}$ is varied between 0.0 and 1.0 at increments of 0.1, and used to produce a weighted mean prediction of interfacial values from respective bulk values. In Figure S5, the absolute percent error (APE) of weighted mean predictions for each unique interfacial IFC2 element are plotted across the full weight scan, organized by the type of IFC2 tensor. Note that a slightly different notation is used, where $w_{GaAs}=w_{mix}$. Points corresponding to interfacial IFC2 elements excluded as outliers based on the above criterion are highlighted in red, and can be seen to produce the highest APE values. Figure S6 plots total MAPE values across the weight scan (and the harmonic mean) for different NN bins and for all terms, with outliers included (Fig. S6a) and with outliers excluded (Figure S6b). The MAPE for all terms, but with outliers excluded, is used to identify the optimal weight, at $w_{mix}=0.5$, which is coincidentally the arithmetic mean. Also plotted are the MAE values in Fig S6c, which are almost entirely unchanged whether outliers are included or excluded.

Figure S7 presents the same kind of bar plot relating bulk and interfacial IFC2s for 3NN terms as that seen in Figure 2 of the main text (for 1NN and 2NN terms). In Figure S8, the sign and absolute value of each interfacial IFC2 element is plotted along with the sign and absolute values of their corresponding bulk values. This is another demonstration of the complex relationship between bulk and interfacial IFCs, as it is evident that the interfacial values can sometimes take on different signs or be nearly and order of magnitude different from their respective bulk values. Also visible are the seven IFC2 elements whose interfacial values are below 0.01 eV/Å; all except the last one (3NN_z3, plane 3, xz) are treated as outliers.

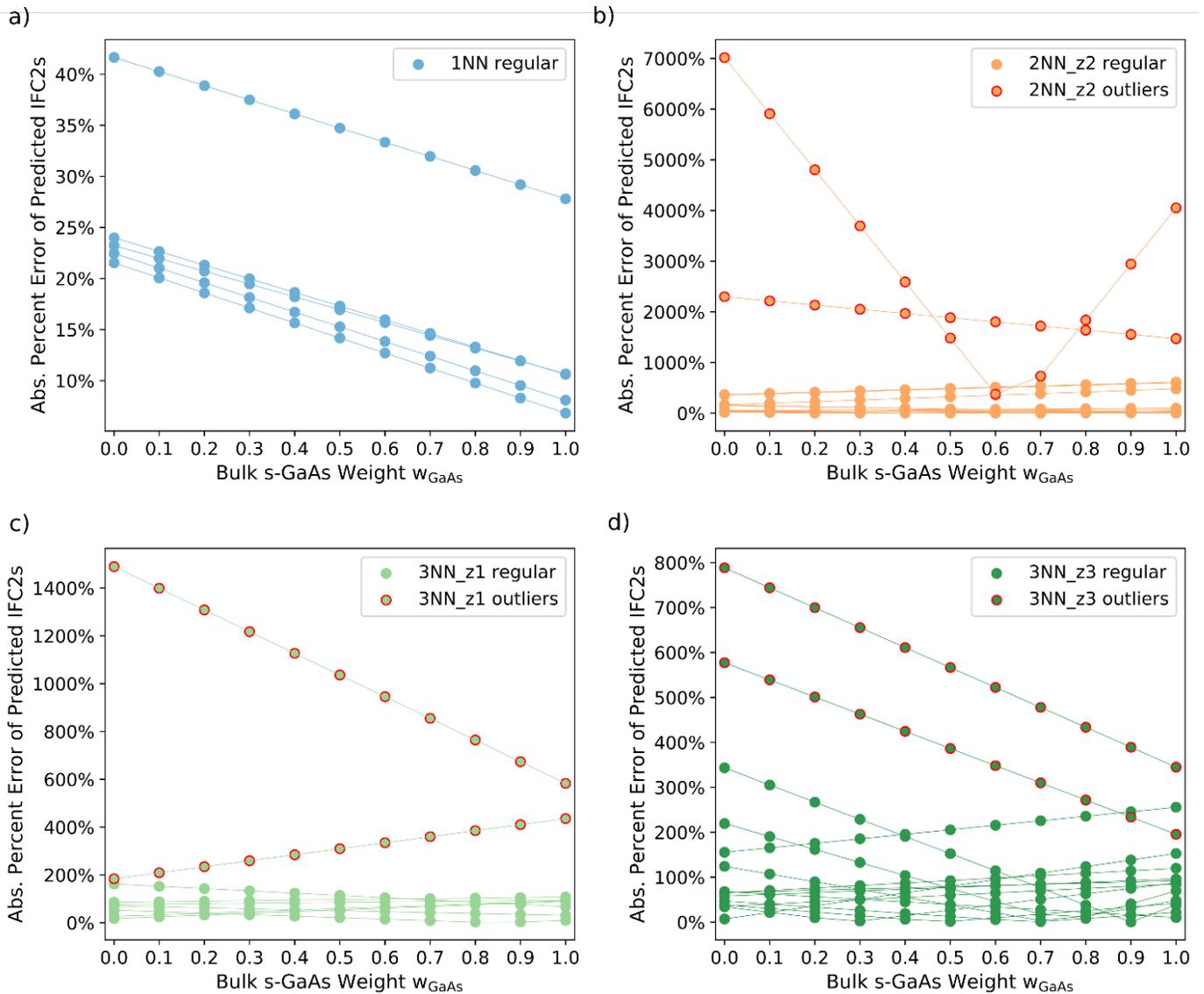

*Figure S5. Absolute percent errors for each interfacial IFC2s predicted with the weighted means of bulk values across different weights, labeled as $w_{GaAs}$ but equivalent to $w_{mix}$, for (a) 1NN, (b) 2NN_z2, (c) 3NN_z1, (d) 3NN_z3. Specific interfacial IFC2s are not*

labeled. However, IFC2s considered outliers are highlighted with red marker edge colors. Note that each plot is plotted on its own scale along the y-axis.

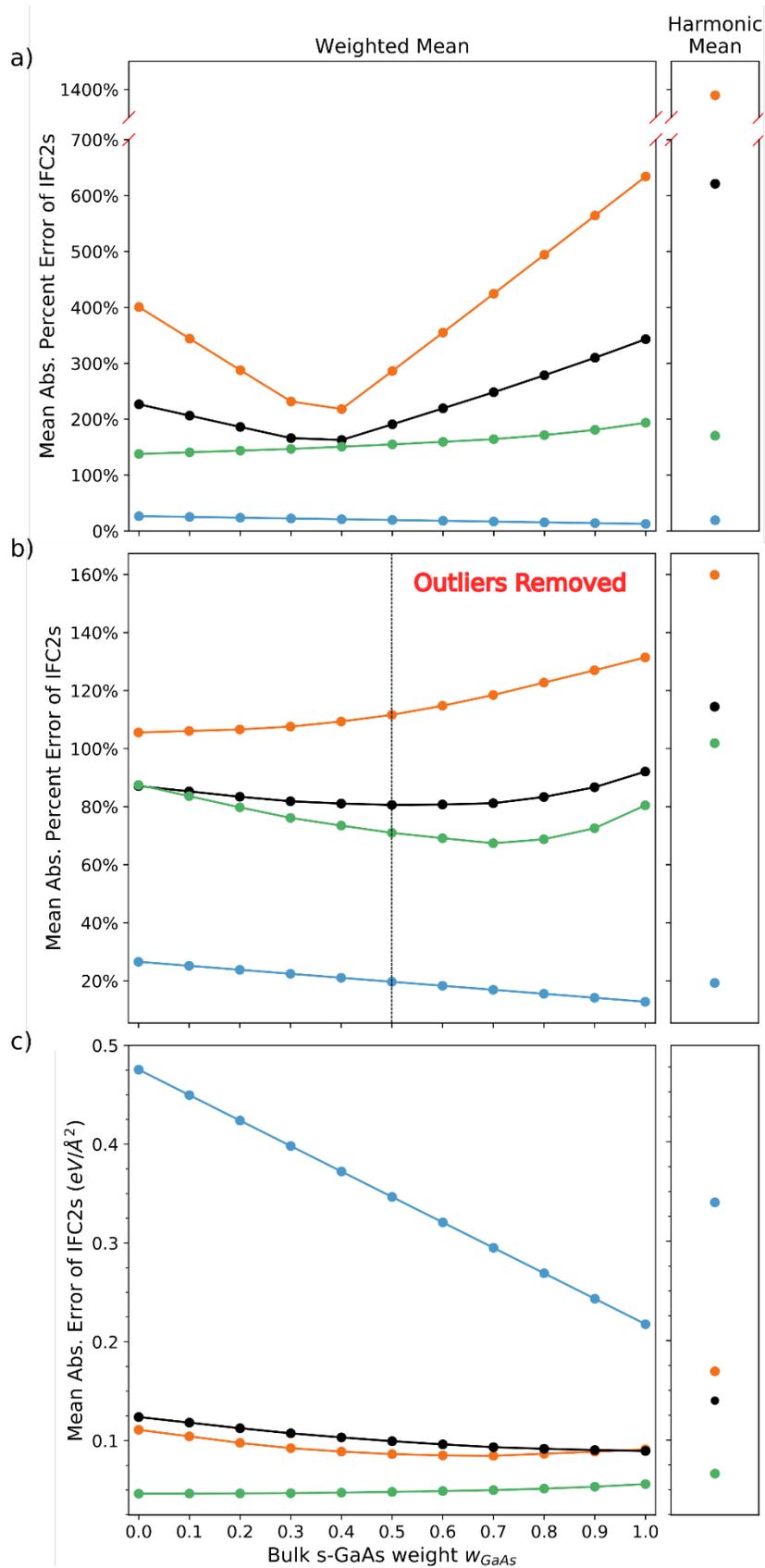

*Figure S6.* Mean absolute percent error for different NN bins and for all values, across different weighted means and for the harmonic mean, are plotted with (a) outliers included and (b) with outliers excluded. Note the break in the y-axis in plot (a). The dashed black

line in (b) indicates the weight that minimizes MAPE across all interfacial IFC2s (outliers excluded), which is coincidently the arithmetic mean. Mean absolute error if plotted in (c) for weighted means and the harmonic mean. Outliers were excluded in generating the plot; however, the MAE plot generated with outliers included is identical.

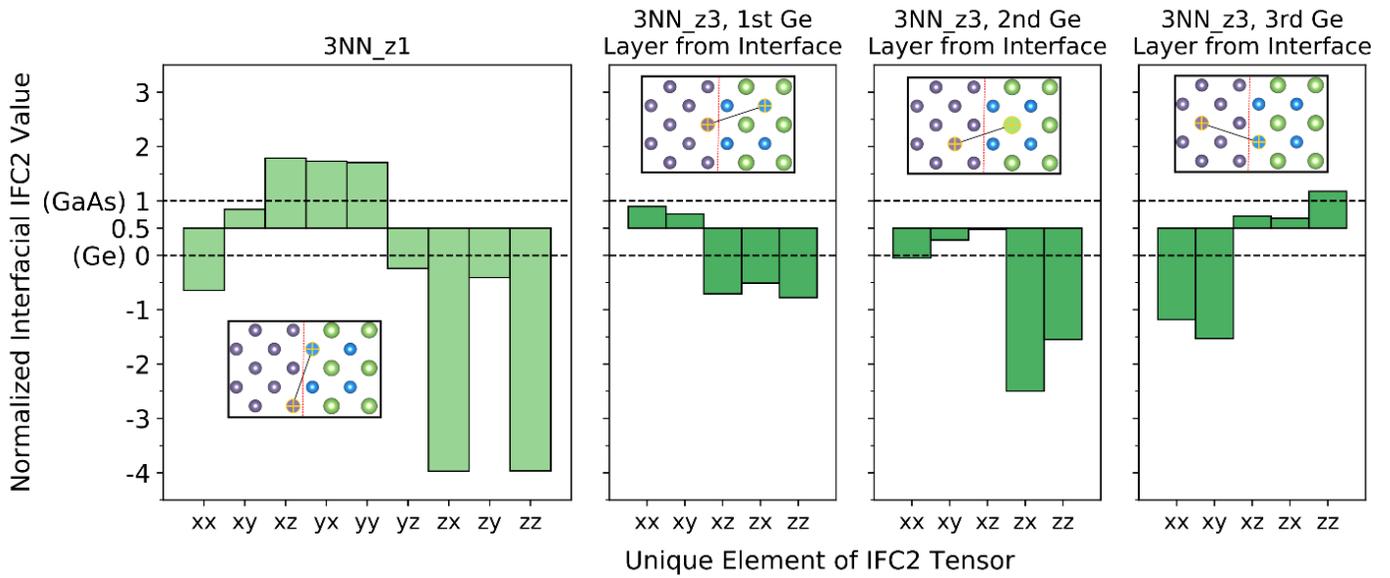

Figure S7. Normalized values of unique IFC2 elements, using Eq. 2 of the main text, for the 3NN interfacial IFC2 tensors. Insets provide representative atom pairs corresponding to each IFC2 tensor.

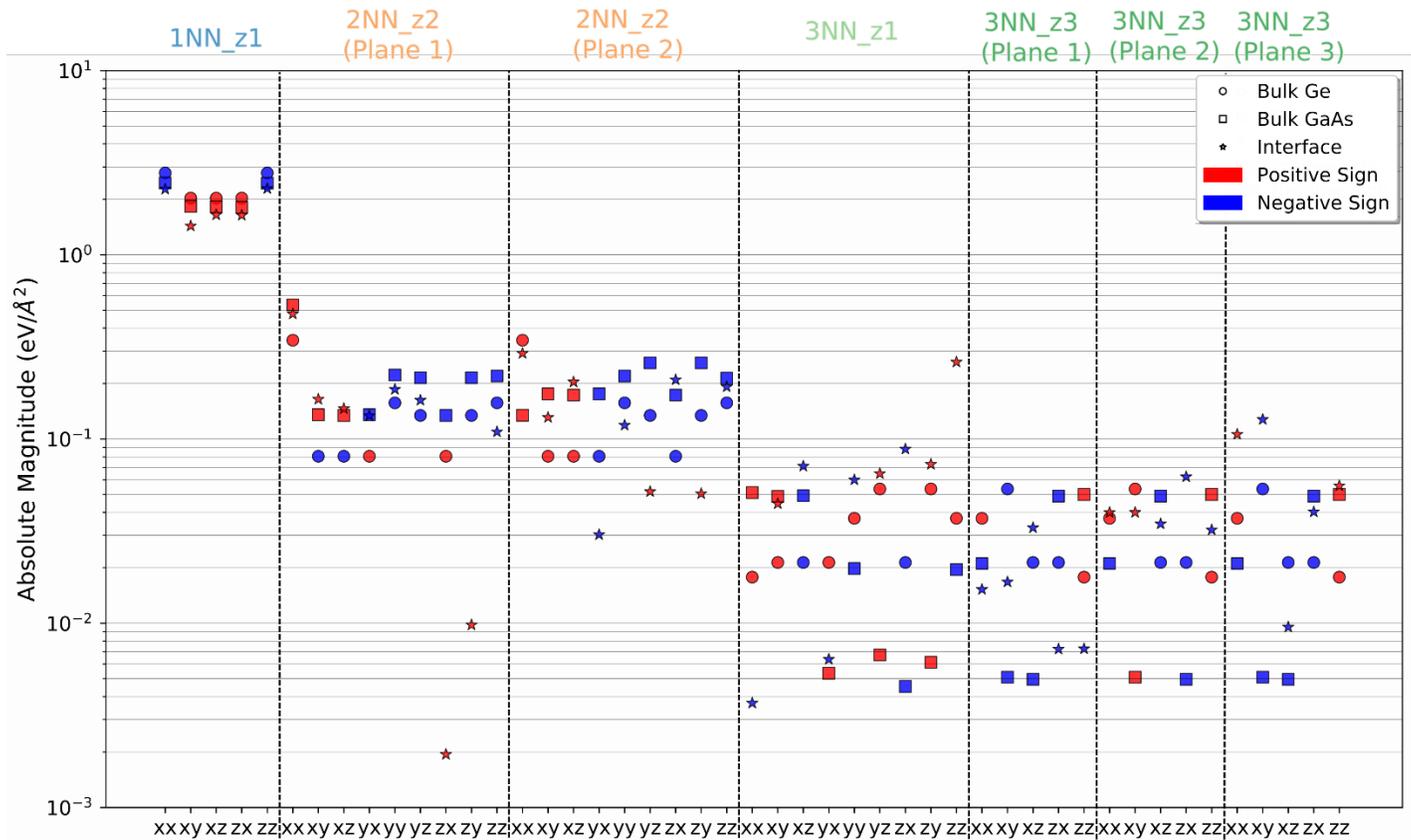

Figure S8. Plot of the absolute magnitude of interfacial IFC2 values and respective bulk values for all unique interfacial IFC2s. Black dashed lines delineate different types of IFC2 tensors, indicated by the labels at the top, while x-axis labels indicate which tensor element each triplet of bulk and interfacial values correspond to. Sign is indicated by color

# Section SIV: Model Selection and Force Performance of Potentials

Model selection was chosen based on stability considerations and on force performance on a holdout validation set. Two primary SNAP hyperparameters were varied: the cutoff of SNAP (between ~3.2 Å and 6.5 Å), and the number of parameters of SNAP, controlled by the hyperparameter $J_{max}$ (=2,3,4). In LAMMPS [11], which implements SNAP, this is controlled by the parameter twojmax, which is literally $2*J_{max}$. For all fits, the element weights used were 1.0, 0.5, and 1.5 for Ge, Ga, and As, respectively; likewise, all element radii were 1.0 and $\theta_{max}$ was set to the default value of 0.99363.

Force errors were calculated for bulk and interface systems on the holdout validation set, using mean percent error (MPE) of the predicted force vector as the primary metric, defined as:

$$MPE = \frac{1}{N} \sum_i \frac{\|F_i - F_i^{ref}\|}{\|F_i^{ref}\|}, \qquad (S2)$$

where N is the number of force vectors and the standard Euclidean norm is used. Stability is assessed by performing 1.5 ns 400K NVE simulations on the bulk Ge, bulk GaAs, and Ge/GaAs DFT supercells and examining whether a system maintains structural integrity and a windowed average temperature (100 timesteps) within 5% of the target temperature of 400K.

Results of these hyperparameters scans are provided in Tables S4-S9 for both uncorrected and corrected versions of pure SNAP fits, anharmonic SNAP + TITEP fits train on the full fitting data, and the anharmonic SNAP + TITEP fits train on just random displacement data. In addition to force MPE, force error is also presented in terms of mean absolute error of force components. Stability is described as either a green check mark, indicating the given potential for a given system is stable, or a red "x" mark indicating the potential was unstable for the given system. Stability for the "all" category simply indicates whether stability was achieved for all systems or not. The fits highlighted in yellow in Table S4 and Table S8 correspond to the final chosen p-SNAP and aS+T potentials, respectively.

Figure S9 plots the test force errors of the chosen p-SNAP and aS+T potential in three different ways: predicted vs target force components (Fig. S9(a)), residual error vs. target force component (Fig S9(b)), and a histogram of absolute force component errors for each potential (Fig S9(c)). Note that while the mean absolute error is higher for aS+T (48.4 meV/Å) than for p-SNAP (45.9 meV/Å), the median force error is actually slightly lower for aS+T (29.5 meV/Å) than for p-SNAP (33.8 meV/Å). Furthermore, there is a higher concentration of low force errors for aS+T compared to p-SNAP fit visible in the histogram of Fig S9(c), while in contrast Fig S9(a-b) suggests that aS+T exhibits higher maximum force errors than p-SNAP. Taken together, this suggests that the force errors of aS+T are generally lower than p-SNAP, but exhibit a longer tail of high force errors that drives up its overall MAE.

Table S4. Performance of different fits of pure SNAP, fit to all fitting data, and with no TITEP correction applied. Highlighted row is final chosen p-SNAP potential.

| Pure SNAP, All Fitting Data, Uncorrected | | | | | | | | |
|---|---|---|---|---|---|---|---|---|
| Twojmax | Rcut (Å) | Bulk Ge | | Bulk s-GaAs | | Ge-GaAs Interface | | All | |
| | | Force MPE (MAE eV/Å) | Stability | Force MPE (MAE eV/Å) | Stability | Force MPE (MAE eV/Å) | Stability | Force MPE (MAE eV/Å) | Stability |
| 4 | 3.217 | 0.11954 (57.47) | ✓ | 0.12995 (60.03) | ✓ | 0.14696 (65.14) | ✓ | 0.13571 (61.90) | ✓ |
| 4 | 4.341 | 0.09778 (48.15) | ✓ | 0.15782 (72.41) | ✓ | 0.15045 (67.31) | ✓ | 0.13910 (63.80) | ✓ |
| 4 | 5.133 | 0.14034 (72.74) | ✓ | 0.14701 (69.05) | ✓ | 0.20211 (90.96) | ✗ | 0.17246 (80.76) | ✗ |
| 4 | 5.907 | 0.11480 (59.29) | ✓ | 0.14726 (71.52) | ✓ | 0.18604 (85.46) | ✗ | 0.15819 (75.31) | ✗ |
| 6 | 3.217 | 0.10139 (49.21) | ✓ | 0.11700 (54.08) | ✓ | 0.13179 (58.22) | ✓ | 0.12036 (54.89) | ✓ |
| 6 | 4.341 | 0.07912 (38.72) | ✓ | 0.10350 (48.47) | ✓ | 0.11809 (52.94) | ✓ | 0.10456 (48.22) | ✓ |
| 6 | 5.133 | 0.08710 (44.11) | ✓ | 0.10159 (47.82) | ✓ | 0.12517 (56.99) | ✗ | 0.10956 (51.40) | ✗ |
| 6 | 5.907 | 0.09122 (45.83) | ✓ | 0.09716 (46.57) | ✓ | 0.12937 (59.90) | ✗ | 0.11152 (52.95) | ✗ |
| 6 | 6.538 | 0.09677 (49.26) | ✓ | 0.12202 (58.17) | ✓ | 0.15996 (73.19) | ✗ | 0.13436 (63.33) | ✗ |
| 8 | 3.217 | 0.08629 (41.90) | ✓ | 0.10479 (48.55) | ✓ | 0.12208 (53.47) | ✓ | 0.10866 (49.30) | ✓ |
| 8 | 4.341 | 0.07175 (35.17) | ✓ | 0.09211 (43.06) | ✓ | 0.10690 (48.05) | ✓ | 0.09427 (43.53) | ✓ |
| **8** | **5.133** | **0.06727 (33.71)** | **✓** | **0.08396 (39.87)** | **✓** | **0.09983 (45.70)** | **✓** | **0.08758 (41.19)** | **✓** |
| 8 | 5.907 | 0.07316 (36.91) | ✓ | 0.07463 (35.52) | ✓ | 0.10228 (47.23) | ✓ | 0.08787 (41.64) | ✓ |

Table S5. Performance of different fits of pure SNAP, fit to all fitting data, and with a TITEP correction applied.

| Pure SNAP, All Fitting Data, Corrected | | | | | | | | |
|---|---|---|---|---|---|---|---|---|
| Twojmax | Rcut (Å) | Bulk Ge | | Bulk s-GaAs | | Ge-GaAs Interface | | All | |
| | | Force MPE (MAE eV/Å) | Stability | Force MPE (MAE eV/Å) | Stability | Force MPE (MAE eV/Å) | Stability | Force MPE (MAE eV/Å) | Stability |
| 4 | 3.217 | 0.03316 (18.67) | ✓ | 0.04110 (22.92) | ✗ | 0.12157 (55.05) | ✗ | 0.07872 (37.67) | ✗ |
| 4 | 4.341 | 0.03889 (22.11) | ✓ | 0.04054 (21.75) | ✓ | 0.11586 (53.09) | ✗ | 0.07721 (37.27) | ✗ |
| 4 | 5.133 | 0.07442 (44.21) | ✓ | 0.07302 (38.75) | ✓ | 0.17760 (80.42) | ✗ | 0.12486 (60.64) | ✗ |
| 4 | 5.907 | 0.06611 (40.23) | ✓ | 0.08516 (48.07) | ✓ | 0.17150 (81.13) | ✗ | 0.12288 (62.38) | ✗ |
| 6 | 3.217 | 0.03187 (17.22) | ✓ | 0.03820 (20.18) | ✓ | 0.12057 (53.80) | ✗ | 0.07717 (35.99) | ✗ |
| 6 | 4.341 | 0.03215 (17.13) | ✓ | 0.04209 (22.29) | ✓ | 0.11062 (50.63) | ✗ | 0.07333 (34.95) | ✗ |
| 6 | 5.133 | 0.05076 (31.45) | ✓ | 0.05071 (27.77) | ✓ | 0.12042 (57.25) | ✗ | 0.08505 (43.21) | ✗ |
| 6 | 5.907 | 0.05031 (31.33) | ✓ | 0.05952 (30.96) | ✓ | 0.12607 (59.98) | ✗ | 0.08997 (45.34) | ✗ |
| 6 | 6.538 | 0.06709 (38.92) | ✓ | 0.07305 (39.06) | ✓ | 0.14313 (67.23) | ✗ | 0.10605 (52.89) | ✗ |
| 8 | 3.217 | 0.02773 (14.73) | ✓ | 0.03653 (19.57) | ✓ | 0.11979 (53.18) | ✗ | 0.07531 (34.90) | ✗ |
| 8 | 4.341 | 0.03350 (17.61) | ✓ | 0.04381 (22.90) | ✓ | 0.10490 (47.95) | ✗ | 0.07130 (33.91) | ✗ |
| 8 | 5.133 | 0.03825 (21.43) | ✓ | 0.04868 (25.51) | ✓ | 0.10293 (48.15) | ✗ | 0.07277 (35.63) | ✗ |
| 8 | 5.907 | 0.04826 (28.76) | ✓ | 0.04703 (24.66) | ✓ | 0.10728 (51.36) | ✗ | 0.07700 (38.84) | ✗ |

Table S6. Performance of different fits of aS+T, fit to all fitting data, and with no correction applied to the TITEP portion

| Anharmonic SNAP + TITEP, All Fitting Data, Uncorrected | | | | | | | | | |
|---|---|---|---|---|---|---|---|---|---|
| Twojmax | Rcut (Å) | Bulk Ge | | Bulk s-GaAs | | Ge-GaAs Interface | | All | |
| | | Force MPE (MAE eV/Å) | Stability | Force MPE (MAE eV/Å) | Stability | Force MPE (MAE eV/Å) | Stability | Force MPE (MAE eV/Å) | Stability |
| 4 | 3.217 | 0.07506 (41.42) | ✓ | 0.07733 (41.07) | ✓ | 0.13873 (66.29) | ✗ | 0.10699 (53.58) | ✗ |
| 4 | 4.341 | 0.10735 (56.17) | ✓ | 0.09622 (49.94) | ✓ | 0.16175 (76.57) | ✓ | 0.13128 (64.61) | ✓ |
| 4 | 5.133 | 0.13811 (77.47) | ✓ | 0.15671 (80.39) | ✓ | 0.21448 (103.05) | ✗ | 0.18048 (90.81) | ✗ |
| 4 | 5.907 | 0.16776 (96.45) | ✓ | 0.19447 (106.55) | ✓ | 0.25350 (124.04) | ✗ | 0.21682 (112.62) | ✗ |
| 6 | 3.217 | 0.05296 (28.56) | ✓ | 0.06069 (31.24) | ✓ | 0.12498 (58.46) | ✗ | 0.09040 (43.97) | ✗ |
| 6 | 4.341 | 0.05964 (31.69) | ✓ | 0.06567 (33.71) | ✓ | 0.12598 (58.77) | ✗ | 0.09385 (45.54) | ✗ |
| 6 | 5.133 | 0.07724 (41.31) | ✓ | 0.07832 (39.53) | ✓ | 0.13288 (62.41) | ✗ | 0.10491 (51.24) | ✗ |
| 6 | 5.907 | 0.09843 (54.62) | ✓ | 0.11375 (57.44) | ✓ | 0.16612 (78.77) | ✗ | 0.13568 (67.24) | ✗ |
| 6 | 6.538 | 0.13144 (73.32) | ✓ | 0.12760 (65.27) | ✓ | 0.19748 (93.86) | ✗ | 0.16297 (81.37) | ✗ |
| 8 | 3.217 | 0.04965 (26.16) | ✓ | 0.05542 (28.01) | ✓ | 0.11998 (55.61) | ✗ | 0.08575 (41.13) | ✗ |
| 8 | 4.341 | 0.05697 (29.83) | ✓ | 0.05791 (29.35) | ✓ | 0.11966 (55.43) | ✗ | 0.08808 (42.31) | ✗ |
| 8 | 5.133 | 0.05600 (29.57) | ✓ | 0.06060 (30.72) | ✓ | 0.11178 (51.93) | ✗ | 0.08464 (40.87) | ✗ |
| 8 | 5.907 | 0.06086 (32.23) | ✓ | 0.06955 (34.74) | ✓ | 0.11997 (55.79) | ✗ | 0.09219 (44.47) | ✗ |

Table S7. Performance of different fits of aS+T, fit to all fitting data, and **with** a correction applied to the TITEP portion

| Anharmonic SNAP + TITEP, All Fitting Data, Corrected | | | | | | | | | |
|---|---|---|---|---|---|---|---|---|---|
| Twojmax | Rcut (Å) | Bulk Ge | | Bulk s-GaAs | | Ge-GaAs Interface | | All | |
| | | Force MPE (MAE eV/Å) | Stability | Force MPE (MAE eV/Å) | Stability | Force MPE (MAE eV/Å) | Stability | Force MPE (MAE eV/Å) | Stability |
| 4 | 3.217 | 0.07261 (40.89) | ✓ | 0.07704 (41.75) | ✓ | 0.13971 (67.52) | ✗ | 0.10678 (54.22) | ✗ |
| 4 | 4.341 | 0.09894 (60.97) | ✓ | 0.08323 (46.35) | ✓ | 0.15612 (76.58) | ✓ | 0.12306 (64.91) | ✓ |
| 4 | 5.133 | 0.12697 (77.38) | ✓ | 0.13426 (76.08) | ✓ | 0.20303 (101.49) | ✗ | 0.16629 (88.92) | ✗ |
| 4 | 5.907 | 0.12909 (82.17) | ✓ | 0.15489 (93.32) | ✓ | 0.23224 (116.37) | ✗ | 0.18649 (101.87) | ✗ |
| 6 | 3.217 | 0.05440 (29.84) | ✓ | 0.06047 (31.86) | ✓ | 0.12626 (59.25) | ✗ | 0.09134 (44.84) | ✗ |
| 6 | 4.341 | 0.05766 (31.40) | ✓ | 0.06206 (32.72) | ✓ | 0.13273 (62.41) | ✗ | 0.09575 (47.01) | ✗ |
| 6 | 5.133 | 0.07132 (39.16) | ✓ | 0.07012 (36.80) | ✓ | 0.13851 (65.18) | ✗ | 0.10409 (51.37) | ✗ |
| 6 | 5.907 | 0.08822 (52.73) | ✓ | 0.09779 (53.20) | ✓ | 0.16789 (80.25) | ✗ | 0.12990 (66.40) | ✗ |
| 6 | 6.538 | 0.10237 (62.27) | ✓ | 0.10109 (56.15) | ✓ | 0.18062 (86.84) | ✗ | 0.14057 (72.80) | ✗ |
| 8 | 3.217 | 0.04865 (25.79) | ✓ | 0.05384 (27.59) | ✓ | 0.12219 (56.97) | ✗ | 0.08619 (41.60) | ✗ |
| 8 | 4.341 | 0.05397 (29.14) | ✓ | 0.05529 (28.59) | ✓ | 0.13033 (61.06) | ✗ | 0.09190 (44.72) | ✗ |
| 8 | 5.133 | 0.05221 (28.06) | ✓ | 0.05423 (28.34) | ✓ | 0.12201 (57.19) | ✗ | 0.08709 (42.48) | ✗ |

| 8 | 5.907 | 0.05653 (31.44) | ✓ | 0.06258 (33.58) | ✓ | 0.12424 (58.49) | ✗ | 0.09142 (45.31) | ✗ |

Table S8. Performance of different fits of aS + T, fit to only random displacement data, and with no correction applied to the TITEP portion of the potential. Highlighted row is final chosen aS+T potential.

| *Anharmonic SNAP + TITEP, Only Random Displacement Data, Uncorrected* | | | | | | | | | |
|---|---|---|---|---|---|---|---|---|---|
| Twojmax | Rcut (Å) | **Bulk Ge** | | **Bulk s-GaAs** | | **Ge-GaAs Interface** | | **All** | |
| | | **Force MPE (MAE eV/Å)** | **Stability** | **Force MPE (MAE eV/Å)** | **Stability** | **Force MPE (MAE eV/Å)** | **Stability** | **Force MPE (MAE eV/Å)** | **Stability** |
| 4 | 5.133 | 0.15545 (83.19) | ✗ | 0.16306 (82.59) | ✓ | 0.23230 (110.47) | ✗ | 0.19524 (96.47) | ✗ |
| 6 | 5.133 | 0.08144 (43.75) | ✓ | 0.08136 (41.55) | ✓ | 0.13914 (65.97) | ✓ | 0.10983 (54.13) | ✓ |
| 8 | 5.133 | 0.05593 (29.95) | ✓ | 0.06550 (33.38) | ✓ | 0.11367 (53.51) | ✓ | 0.08681 (42.43) | ✓ |

Table S9. Performance of different fits of aS+T, fit to only random displacement data, and with a correction applied to the TITEP portion of the potential

| *Anharmonic SNAP + TITEP, Only Random Displacement Data, Corrected* | | | | | | | | | |
|---|---|---|---|---|---|---|---|---|---|
| Twojmax | Rcut (Å) | **Bulk Ge** | | **Bulk s-GaAs** | | **Ge-GaAs Interface** | | **All** | |
| | | **Force MPE (MAE eV/Å)** | **Stability** | **Force MPE (MAE eV/Å)** | **Stability** | **Force MPE (MAE eV/Å)** | **Stability** | **Force MPE (MAE eV/Å)** | **Stability** |
| 4 | 5.133 | 0.14059 (79.83) | ✓ | 0.15500 (82.98) | ✓ | 0.22677 (111.19) | ✗ | 0.18672 (96.08) | ✗ |
| 6 | 5.133 | 0.08179 (45.18) | ✓ | 0.07619 (40.19) | ✓ | 0.14385 (68.65) | ✗ | 0.11091 (55.46) | ✗ |
| 8 | 5.133 | 0.05329 (29.18) | ✓ | 0.06039 (31.85) | ✓ | 0.11855 (56.06) | ✗ | 0.08724 (43.10) | ✗ |

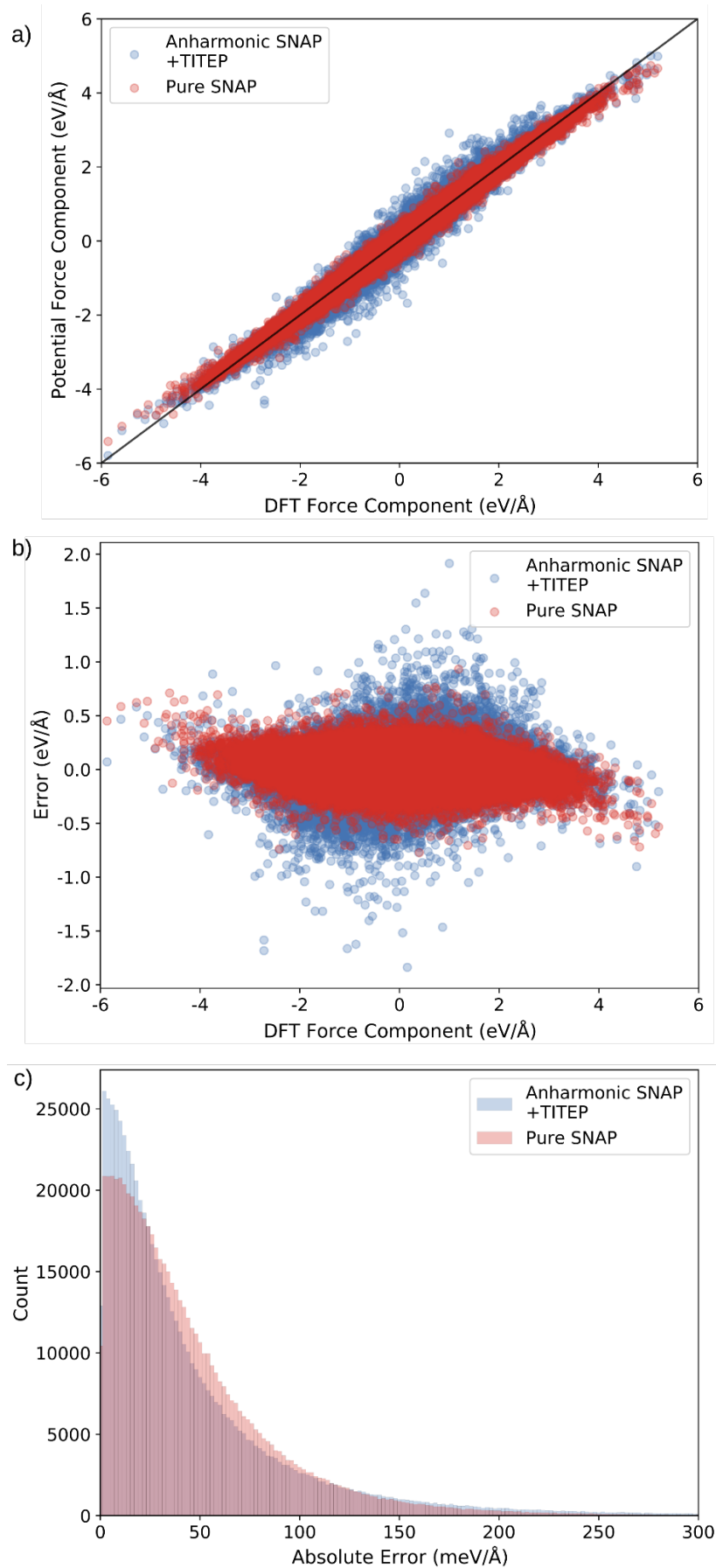

*Figure S9. Plots of (a) predicted vs reference DFT force components, (b) residual error vs. reference DFT force component, and (c) histograms of absolute force component errors for both pSNAP and aS+T.*

# Section SV: Thermal Properties of Fitted Potentials

Alongside Figure 4 of the main text, Figure S10-S13 are symmetric log-log plots comparing predicted and reference IFC2 and IFC3. These plots enable both positive and negative-valued IFCs to be plotted on a log scale. To avoid a divergence near zero, the region between $-10^4$ and $10^4$ eV/Å² on both the x and y axis are plotted linearly rather than logarithmically. In Figure S10 and Figure S11, which plot the relationship between predicted and reference IFC2s for bulk Ge and bulk s-GaAs respectively, the p-SNAP potential predicts a number of 4NN+ IFC2s to be zero (or very close to it). This is a consequence of the limited cutoff of the SNAP potential, 5.133 Å, as many of those IFC2s correspond to atom pairs separated by distances much larger than the cutoff. In contrast, the TITEP portion of the aS+T potential includes the full range of IFC2s in these bulk systems. While beyond the scope of this paper, a possible fruitful path for future researchers would be to better understand the relationship between how well specific IFC2 and IFC3s are reproduced by given potentials (as displayed in these plots) compared to how well dispersions and thermal conductivities are reproduced.

Figure S14 plots all absolute percent errors of predicted IFC2s for the interface system compared to the absolute magnitude of the reference IFC2 values. This data is summarized in Figure 5 of the main text and provides a more complete picture of the underlying spread of data.

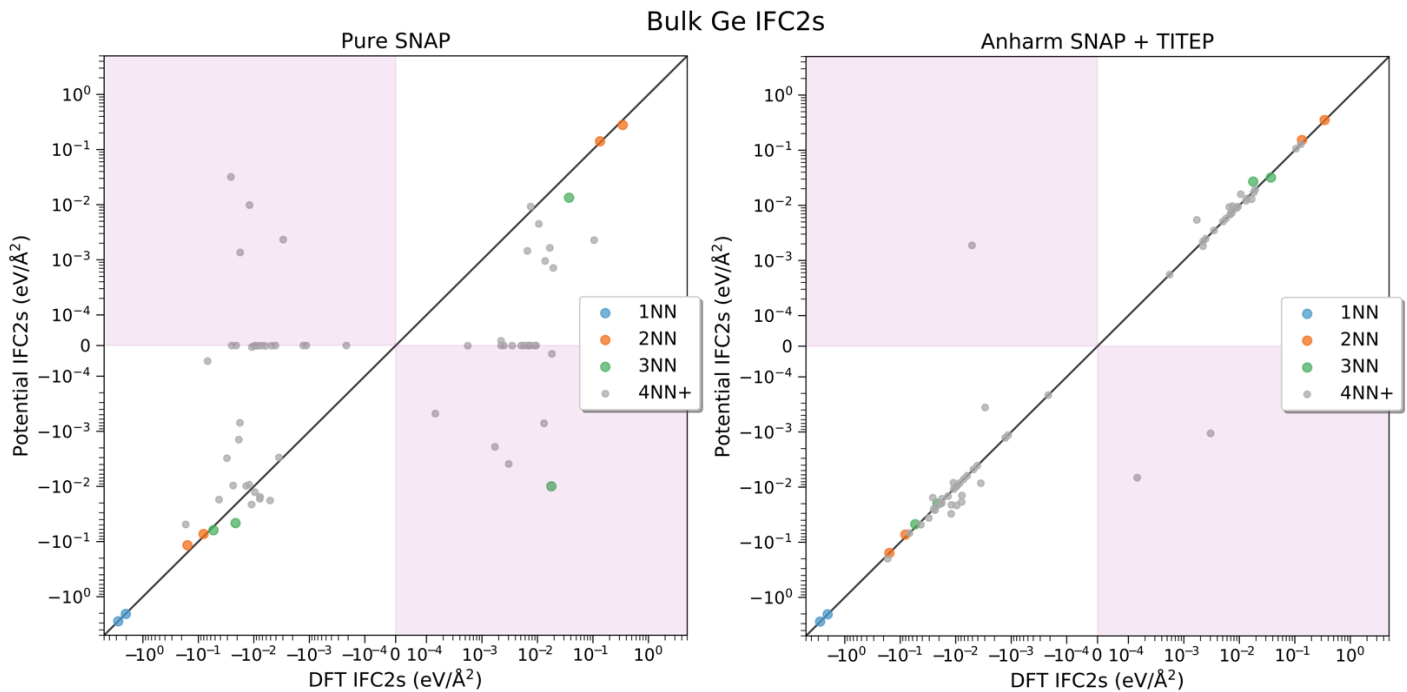

*Figure S10. Symmetric log-log plot of predicted bulk Ge IFC2s against DFT bulk Ge IFC2s for the p-SNAP (left) and aS+T (right) potential. The red shaded region indicates regions of sign flipping.*

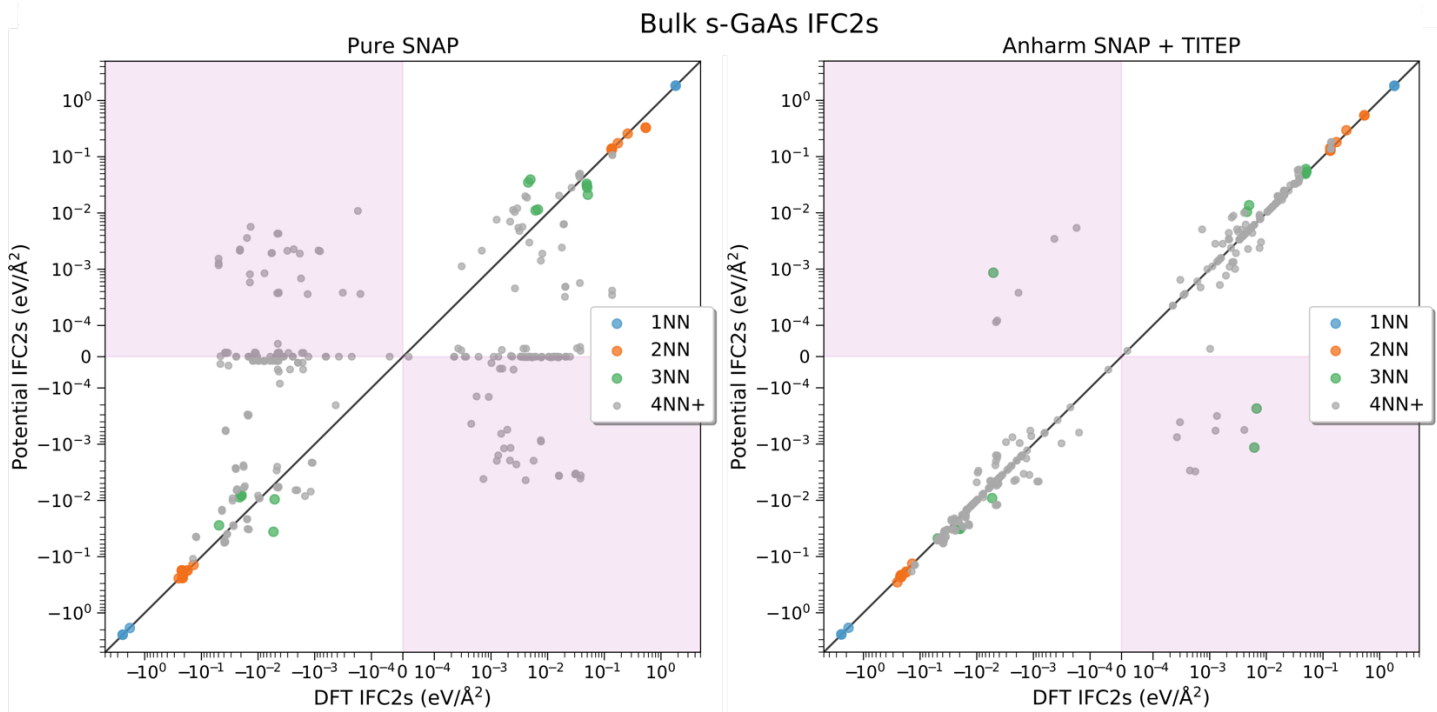

*Figure S11. Symmetric log-log plot of predicted bulk s-GaAs IFC2s against DFT bulk s-GaAs IFC2s for the p-SNAP (left) and aS+T (right) potential. The red shaded region indicates regions of sign flipping.*

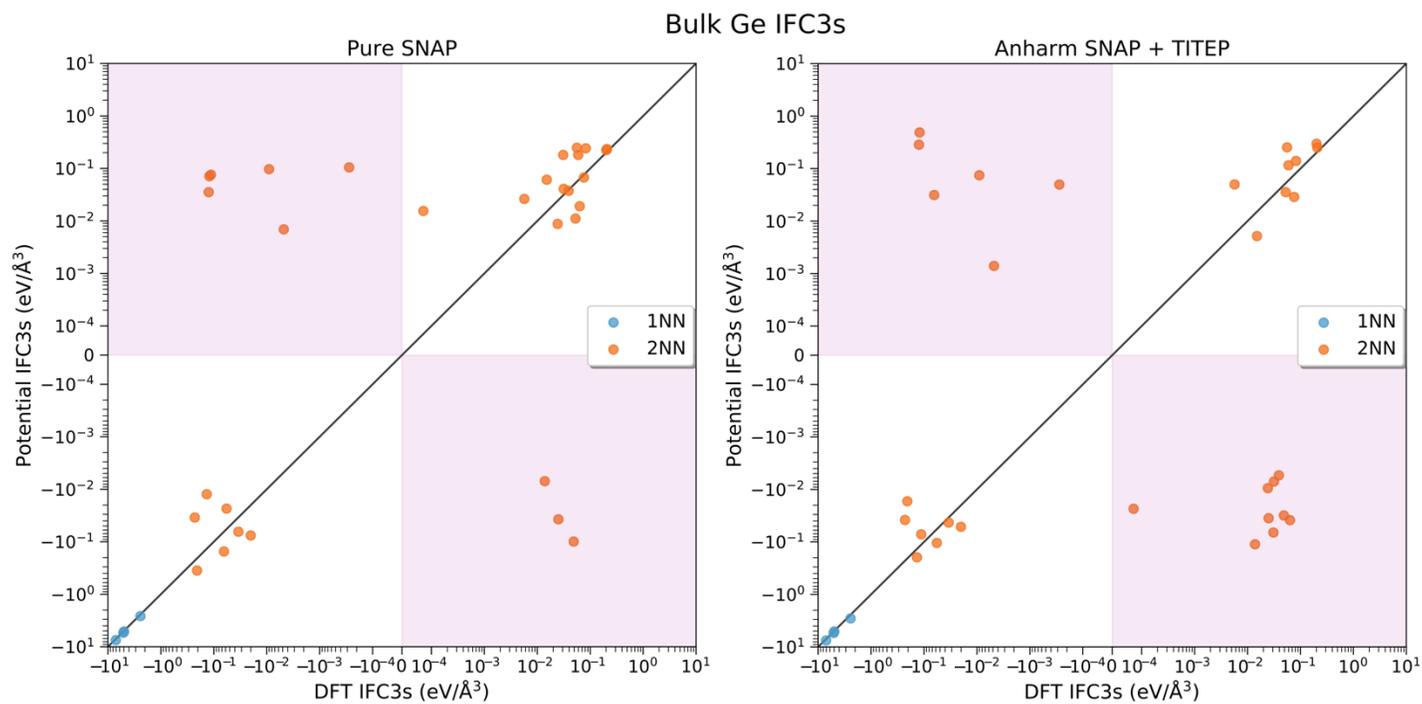

*Figure S12. Symmetric log-log plot of predicted bulk Ge IFC3s against DFT bulk Ge IFC3s for the p-SNAP (left) and aS+T (right) potential. The red shaded region indicates regions of sign flipping.*

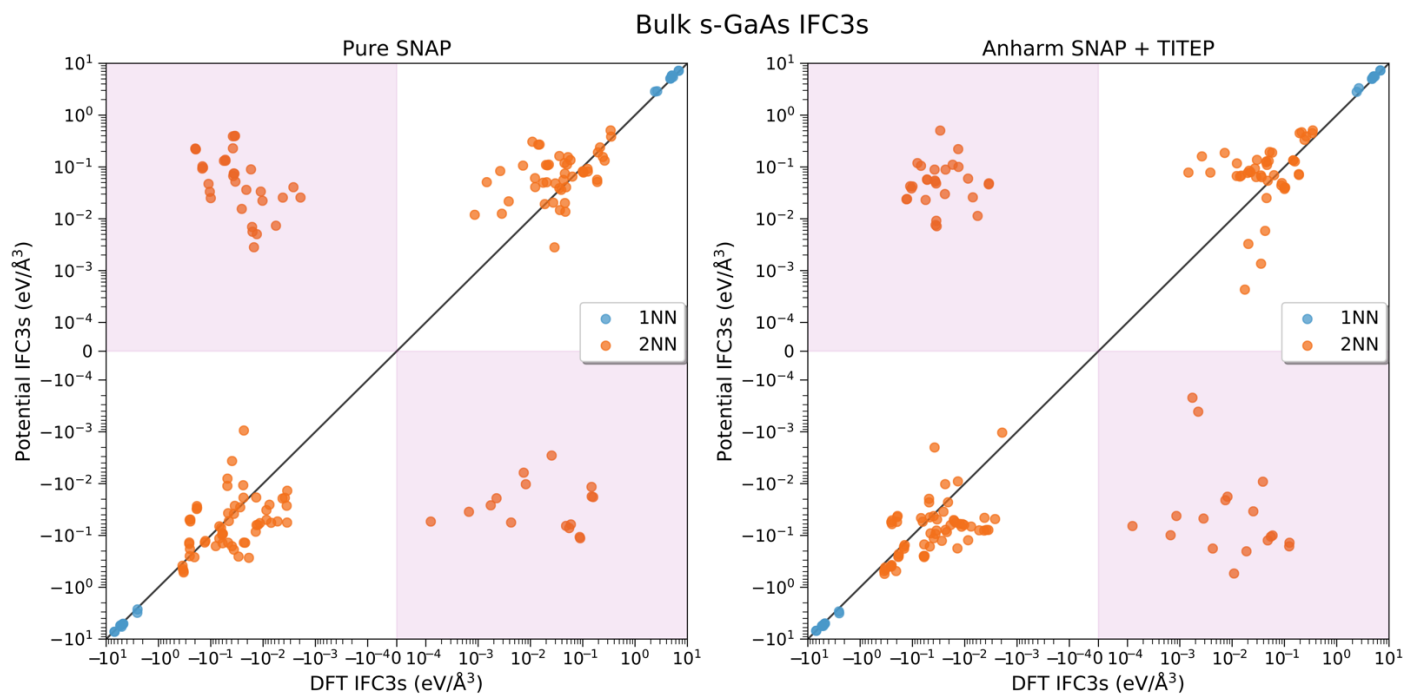

*Figure S13. Symmetric log-log plot of predicted bulk s-GaAs IFC3s against DFT bulk s-GaAs IFC3s for the p-SNAP (left) and aS+T (right) potential. The red shaded region indicates regions of sign flipping.*

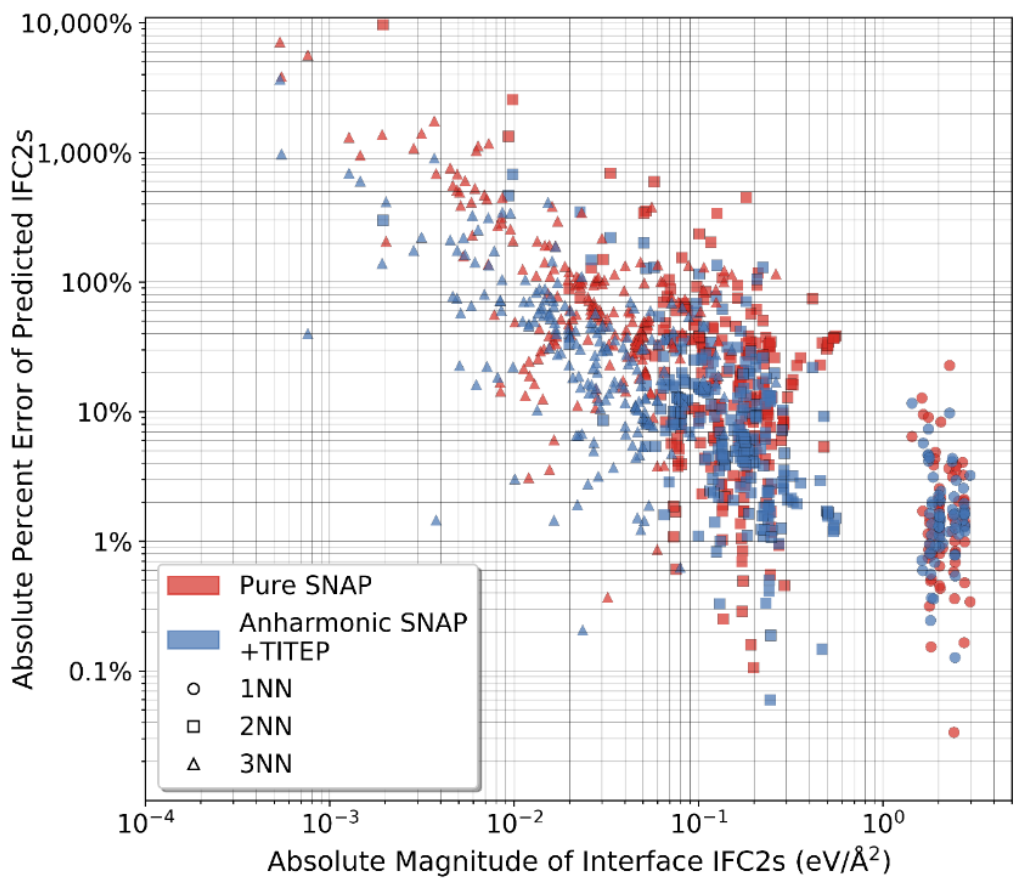

*Figure S14. Log-log plot of all the absolute percent error of unique interface IFC2s predicted by p-SNAP and aS+T against the absolute magnitude of the reference DFT IFC2. Nearest neighbor bin is indicated by shape of marker.*

# Section SVI: Prior, Fully Stable, Anharmonic SNAP + TITEP

As mentioned in the main text, the fully stitched together aS+T potential fitted for this paper was unstable when run in MD, an issue that appears to be due to issues with creating a supercelled version of the interface TITEP and not with the original stitching procedure. Prior to the work of this paper, a preliminary anharmonic SNAP + TITEP potential was generated for Ge/GaAs that, when fully stitched together for a large interface model, was fully stable at 400K and above.

Here, we provide a brief summary of the key features and differences of this preliminary potential. Like the interface described in the main text of this paper, the Ge/GaAs interface of the preliminary potential was modeled in a superlattice geometry, with a pure As-terminated interface for both the middle and periodic boundary interface, and was of a similar length as that described in this paper. However, unlike the interface in this paper, the Ge region was strained to the GaAs lattice parameters in the in-plane direction (rather than vice versa), and the Ge region was half a unit cell larger than the GaAs region (rather than vice versa) to account for the nonstoichiometry necessary to have both interfaces As-terminated. Similar sized bulk supercells systems were used as that described in this paper. Bulk Ge was strained to the GaAs lattice parameters, without relaxing to allow for a compensating tetragonal distortion.

While the *ab initio* data set used to fit the preliminary potential were also generated with projector-augmented, plane-wave DFT using VASP, the exchange and correlation portion of the DFT functional were treated under the generalized gradient approximation (GGA) using the parameterization of Perdew, Burke, Ernzerhof (PBE), instead of the LDA functional used in this paper. Additionally, the plane wave cutoff used was 520 eV, and the k-point meshes used to generate fitting data for bulks and interface systems was 2x2x2 and 2x2x1, respectively. Other electronic parameters were similar to that described in the paper. Bulk and interface IFC2s were generated using Alamode with the small finite displacement method, using ~50 (bulks) or ~200 (interface) randomly displaced structures with displacement magnitude between 0.01 – 0.03 Å as input to Alamode. Unlike in this paper, these small finite displacement calculations were generated less rigorously with the same k-point mesh sizes as that used to generate fitting data.

The preliminary anharmonic SNAP + TITEP was generated using a similar procedure as that described in the main text. Around 900 randomly displaced configurations were used as fitting data, with displacement ranges of 0.05 – 0.1 Å and 0.08 – 0.15 Å (~180 bulk s-Ge, ~180 bulk GaAs, ~540 interface configurations). The key hyperparameters of the anharmonic SNAP were twojmax=8 and a rcut=4.445 Å; all other hyperparameters were the same as that used in this paper. Importantly, the TITEP portions of the potential were corrected such that exact bulk and interface IFC2s were reproduced.

The final potential was stable at 400K for both the individual systems and for a large stitched-together interface system. By virtue of using a corrected version of the anharmonic SNAP + TITEP, the bulk dispersions were exactly reproduced. Thermal conductivity of bulk s-Ge and bulk GaAs were 26% and 24% of their *ab initio* values, respectively. To provide a fair comparison in force error to the potentials described in this paper, a test set of ~300 interface configurations generated by AIMD and sample-TITEP data was created using the preliminary interface structure and PBE electronic parameters. The force MPE and MAE of the preliminary potential against this interface-only test set was 15% and 52.2 meV/Å respectively.